\documentstyle[12pt,aaspp4,epsf]{article}

\def\ltsim{ \,{}^<_\sim\, }
\def\gtsim{ \,{}^>_\sim\, }
\def\etal{{\it et~al.}}
\def\ie{{\it i.e.}}
\def\eg{{\it e.g.}}
\def\cf{{\it cf.}}
\def\hst{{\it HST}}

\lefthead{Harris \etal }
\righthead{Globular Cluster Ages}

\begin{document}

\title{NGC 2419, M92, and the Age Gradient in the Galactic Halo}

\author{William E.~Harris}
\affil{Department of Physics \& Astronomy, 
McMaster University, Hamilton, Ontario L8S 4M1
\\Electronic mail: harris@physics.mcmaster.ca}

\author{Roger A.~Bell}
\affil{University of Maryland, Department of Astronomy,
       College Park MD 20742-2421
\\Electronic mail: rabell@astro.umd.edu}

\author{Don A.~VandenBerg}
\affil{University of Victoria, Department of Physics \& Astronomy,
       PO Box 3055, Victoria, British Columbia  V8W 3P7
\\Electronic mail: davb@uvvm.uvic.ca}

\author{Michael Bolte}
\affil{UCO/Lick Observatory, University of California, Santa Cruz, CA 95064
\\Electronic mail: bolte@lick.ucolick.org}

\author{Peter B.~Stetson, James E.~Hesser, and Sidney van den Bergh}
\affil{National Research Council, Herzberg Institute of
Astrophysics, Dominion Astrophyical Observatory, 5071 West Saanich Road,
Victoria, British Columbia V8X 4M6
\\Electronic mail: firstname.lastname@hia.nrc.ca}

\author{Howard E.~Bond}
\affil{Space Telescope Science Institute, 3700 San Martin Drive,
       Baltimore MD 21218 
\\Electronic mail: bond@stsci.edu}

\author{Gregory G.~Fahlman and Harvey B.~Richer}
\affil{Department of Geophysics \& Astronomy, University of British
       Columbia, Vancouver, British Columbia V6T 1Z4 
\\Electronic mail: lastname@astro.ubc.ca}

\clearpage

\begin{abstract}
The WFPC2 camera on $HST$ has been used to obtain photometry of
the low-metallicity ([Fe/H] $=-2.14$), outer-halo globular cluster
NGC 2419.  Our color-magnitude diagram in $(V, V-I)$ reaches
$V_{lim} \simeq 27.8$, clearly delineating the subgiant and turnoff 
region and about three magnitudes of the unevolved main sequence.
A differential fit of the NGC 2419 CMD to that of the similarly 
metal-poor `standard' cluster M92 shows
that they have virtually identical principal sequences and thus the
same age to within 1 Gyr.  Previously published studies of many
other low-metallicity globular clusters throughout the
Milky Way halo show that they 
possess this same age to within the $\sim 1$ Gyr precision of measurement.  
The addition of the remote-halo object NGC 2419 to this list leads us to
conclude that the earliest star (or globular cluster) formation
began {\it at essentially the same time} everywhere in the Galactic halo 
throughout a region now almost 200 kpc in diameter.
Thus for the metal-poorest clusters in the halo there is no detectable
age gradient with Galactocentric distance.
To estimate the absolute age of NGC 2419 and M92, we fit newly
computed isochrones transformed through model-atmosphere calculations to the
$(M_V,V-I)$ plane, with assumed distance scales that represent the
range currently debated in the literature.  Unconstrained isochrone fits give
$M_V(RR) \simeq 0.55 \pm 0.06$ for both clusters, and a resulting 
age of 14 to 15 Gyr.  Incorporating the full effects of 
helium diffusion would further reduce this 
estimate by $\sim 1$ Gyr.  The first reports of {\it Hipparcos} 
parallax measurements for the lowest-metallicity subdwarfs 
suggest that the distance scale could be as
bright as $M_V(RR) = 0.15$ for [Fe/H] $\simeq -2$, which would require
the cluster ages to be less than 10 Gyr; however, the isochrone fits
for a distance scale this extreme leave several serious
problems which have no obvious solution in the context of 
current stellar models.

\end{abstract}


\clearpage

\section{Introduction}

When did the halo of the Milky Way begin to form, and how long did it
take?  The possible models for halo formation are still bounded by the two classic
extremes of Eggen \etal\ (1962 = ``ELS'', in which the halo stars and 
globular clusters condense out of a rapid, monolithic collapse of the protogalaxy)
and Searle \& Zinn (1978, in which the Galaxy assembles piecemeal over a much
longer period from small, initially independent gas clouds).

The best-explored route to answering these questions has been through the 
age calibration of the Galactic globular clusters (GGCs). Obtaining
accurate and precise measures of their absolute ages, and their dispersion in ages, 
is well known to be a challenging problem which strongly couples both observation
and theory.  Some discussions
assert that the age dispersion is several Gyr, which would
argue strongly against any pure fast-collapse model of formation (\eg\ \cite{cha96};
\cite{sar89}; \cite{sal93}; \cite{lee94}).  
Other studies employing a more selected set of the best available observations 
(\eg\ \cite{ste96} [hereafter SVB]; 
VandenBerg \etal\ 1990, 1996 [hereafter VBS90, VBS96]; \cite{DH93};
\cite{ric96}; \cite{wei97}), favor the interpretation of an age distribution with
a narrow ($< 1$ Gyr) peak and a long, sparsely populated tail to
younger ages, which would be more consistent with an ELS-style collapse.
An important recent development has been the refinement of methods for precise measurement
of age {\it differences} among clusters independently of distance and reddening
(VBS90; VBS96; Sarajedini \& Demarque 1990; \cite{cha96}).
With high-precision photometry at its current limits, it is now possible
to determine {\it relative} cluster ages to within $\pm 0.5$ Gyr, and even
narrower limits ultimately lie within reach (see SVB).

The interpretive models continue to evolve with the data.  \cite{Sa90} has
reformulated the original ELS picture to allow for a spectrum of density fluctuations
within the protohalo.  Conversely, other authors including \cite{Zi93},
\cite{lee93}, and \cite{vdb93} have extended 
the basic Searle-Zinn view to raise the possibility that much
of the halo might have accreted later in the form of 
comparatively few, large dwarf galaxies.  
In addition, it has become increasingly apparent that the Milky Way halo clusters
define clumpy regions within phase space (\eg\ \cite{rog84}; Zinn 1993; \cite{maj94}; 
\cite{van94}; \cite{joh96}; \cite{lyn95}; \cite{fus95}) which may be the
relics of major accretion events.  The Sagittarius system that we now
see being disrupted along with its small retinue of clusters (\eg\ \cite{iba97};
\cite{dac95}) would then be only the most recent such infall event.  
More direct evidence for several possibly distinct 
epochs of cluster formation can be found in the convincing
demonstrations (\eg, \cite{GO88}; \cite{pbs+89}; \cite{Bo89}; 
\cite{GN90}; Buonanno \etal\ 1990, 1993; \cite{sar97}) 
that at least a few clusters in the
mid-halo region are substantially younger --- sometimes by 
as much as 30\% --- than the mean age of the GGCs.   

A key question capable of strongly influencing the competing
models is the existence or absence of any {\it age gradient} in the Galactic halo:
does cluster age depend clearly and systematically on Galactocentric distance?
\cite{SZ78}, \cite{Zi93}, and Lee \etal\ (1994) build a case based on 
horizontal-branch morphology and cluster kinematics that the clusters formed 
over a progressively longer spread of times at larger
Galactocentric distance, and thus that the mean cluster age should decrease
with increasing $R_{gc}$.  However, truly direct 
age measurements must be obtained through photometry of
deeper levels in the color-magnitude diagram (CMD), \ie\ the turnoff and
unevolved main-sequence regions.  With the \hst\ cameras, age measurements from 
main-sequence photometry are now possible for even the most remote known
Milky Way halo clusters.
The very oldest systems in the halo are, by most available evidence, 
most likely to be the globular clusters of lowest metallicity (\eg, VBS96). 
Thus by a careful study of these clusters, which are found everywhere in the halo,
we may obtain a strong lower limit to the true age of the Galaxy.
Similarly, the age {\it range} among these same low-metallicity clusters
gives us an excellent way to estimate when the different parts of the halo
began star formation.

This paper is the first in a series of \hst\ color-magnitude studies for 
the globular clusters in the outermost halo of the Milky Way.
In this paper, we present an age analysis of the outermost-halo, low-metallicity
cluster NGC 2419 (= C0734+390; $\alpha_{2000} = 7^h 38^m 05\fs5$, 
$\delta_{2000}$ = $+38^o 52' 55''$ ; 
$\ell = 180\fdg 4, b = +25\fdg2 $; $R_{gc}\sim 90$ kpc; see \cite{har96}).
In subsequent papers, we will present CMD analyses for the other five
clusters at $R_{gc} \gtsim 80$ kpc (Palomar 3, 4, 14, Eridanus, and AM-1).
Since most of these clusters exhibit the ``second-parameter'' horizontal-branch
anomaly in its most extreme form, in their
totality they will provide a stringent test of the formation scenarios 
mentioned above.

In certain respects, NGC 2419 is arguably the most unusual cluster in the
outer Milky Way halo.  It is much more luminous than the 
other outer-halo clusters, with an absolute
visual magnitude $M_V^t \simeq -9.5$ (\cite{har96})
that places it among the five most luminous clusters in the Galaxy.
Its metallicity of [Fe/H] $\simeq -2.14$ (\cite{zin85}; \cite{sun88})
puts it clearly in the most metal-poor group of known GGCs. But
in contrast to the other outer-halo globulars and most of the dwarf
spheroidal galaxies that inhabit the same region of space, NGC 2419 has
a horizontal branch which is rather uniformly populated from blue to
red like other classic low-metallicity objects such as M15, M92, and M68.  
Thus its HB morphology is {\it not} strikingly unusual for its metal abundance.
However, NGC 2419 cannot simply be interpreted as (for example) a metal-poor
cluster that might have formed initially deep in the inner halo and then migrated out
on a highly elliptical orbit. Its very large core radius and half-mass radius
($r_c \sim 9$ pc, $r_h \sim 19$ pc) are entirely characteristic of the most
remote clusters and unlike any inner-halo object (typically $r_c \sim 1$ pc,
$r_h \sim 3$ pc for clusters at $R_{gc} \sim R_{\odot}$). These features along
with the well known systematic increase of $r_h$ with Galactocentric
distance (\eg\ \cite{van91}; \cite{van95a}) demonstrate that it belongs
to the outermost-halo group as much as any of the other outer-halo clusters.
Since it is at a very different place in the halo from all the other
lowest-metallicity clusters, NGC 2419 holds considerable interest 
for the Galactic age gradient question.

The first color-magnitude study of NGC 2419, by \cite{rac75} from
photographic plates, was barely sufficient to reveal
the nature of the brighter parts of the CMD (the giant branch and horizontal
branch), but fainter features such as the turnoff and subgiant stars were
hopelessly beyond reach of the technology of the time.
The first CCD-based photometry of the cluster (Christian \& Heasley 1988)
reached three magnitudes deeper and thus just barely resolved the turnoff stars.
Their data were sufficient at least
to indicate that NGC 2419 was similar in age to the ``normal'', inner-halo
globular clusters to within a few Gyr. Our new \hst\ photometry,
as will be seen below, reaches considerably deeper still, and now
allows us to carry out an age comparison that is as precise as for any
other cluster in the Milky Way.\footnote{Our study employs data from Cycle 4
programs 5481 and 5672.  Preliminary discussions of this work are given in
\cite{har95}, \cite{hes95}, Hesser \etal\ (1996a,b), and Richer \etal\ (1996).}

In brief, the main purpose of this paper is to estimate the relative age of
NGC 2419 in comparison with normal nearby clusters of
similar metal abundance.  In \S2, we briefly describe the CMD;
in \S3, we estimate its age relative to M92;  
in \S4, we present sample isochrone fits to both NGC 2419 and M92
for a new set of stellar models and theoretical transformations,
under different assumptions for the Population II distance scale;
and in \S5, we discuss briefly the
significance of these results for the early history of the Galaxy.

\section{Color-Magnitude Diagrams}

In Figure \ref{fig1}, we show the composite color-magnitude array (CMD) for
NGC 2419 as derived from our Cycle 4 \hst\ imaging with the WFPC2 camera.
The complete data reduction and calibration are described in \cite{ste97}. 
To isolate a sample of stars which most narrowly defines the cluster sequences
that we are particularly interested in (the lower giant branch, subgiant
branch, and main sequence), we have selected from the \cite{ste97} dataset
all the measured stars
farther than a projected radius of $R = 50''$ from cluster center; inside 
this radius, the photometric scatter increases noticeably because of crowding
and higher background.  We further eliminated individual 
stars with especially uncertain photometry ($\sigma(V), \sigma(V-I)$,
or $\chi$ which stand off by more than three standard deviations from the
mean values at any $V$ magnitude).  This culling procedure left a final 
sample of 17275 stars.

To define the principal sequences of the CMD, we took mean magnitudes
and colors in $0.1-$ or $0.2-$mag bins, with three iterations of
outlier rejection.  For the sparsely populated bright end ($V < 21$) 
of the red-giant branch (RGB) we added the stars from within
$R < 50''$ (not shown in Fig. \ref{fig1}) to help define the RGB locus more accurately.  
The mean lines from the inner region, for the bright stars that are least
affected by crowding ($V \ltsim 24$), are identical with those from the
outer region shown here to well within $\pm0.01$ mag in color at every point;
however, for the fainter main-sequence stars, the data from the inner region
become severely incomplete for $V \gtsim 25$ and we rely on only the mean points
from the outer region to define the faint end of the CMD.
The mean points are  displayed in Figure \ref{fig2} and listed in
Table 1:  in the Table, the first group of entries gives the mean points
for the HB and the remainder give the giant branch and main sequence. 
The number of stars in each bin is listed in the last column.  Clearly, the accuracy
and depth of the \hst\ photometry, combined with the very large sample
of stars, permit the cluster sequences to be defined to a level of precision
that is fully comparable with other, much nearer, clusters that have been
well studied from the ground.  We note in passing that
the zeropoints of the magnitude and color scales are 
tied to the \hst\ ($V,I$) photometric system, but
rely on a preliminary tie-in to ground-based photometry of the same fields and
may therefore contain a residual offset of up to $\pm 0.02$ mag.  This
point should be kept in mind for the later discussion on the
absolute reddening and distance of NGC 2419 (\S4.3 below).  However,
this will not affect the {\it differential} analysis of NGC 2419 relative
to other, ``standard,'' metal-poor clusters, to which we will pin its
age measurement (\S4).

Our new CMD confirms the basic conclusions of the previous studies
(Racine \& Harris 1975; Christian \& Heasley 1988) that NGC 2419
strongly resembles other very metal-poor objects like M92 and M15,
with its steep giant branch and predominantly blue horizontal branch
that extends, with obvious gaps, to very high temperature.
Further detailed discussion of the CMD morphology is given by
\cite{ste97}.  Here, we concentrate on its age analysis.

\section{Relative Age Estimation}

A well known method for precise determination of {\it relative} ages
of globular clusters is to compare the positions of age-sensitive
features in the CMD, such as the color difference between
the main-sequence turnoff (MSTO) and the base of the red-giant branch,
or the magnitude difference $\Delta V(MSTO-HB)$
between the MSTO and horizontal branch (HB).
The color-difference approach was first fully employed by 
VBS90 (also see \cite{sar90}); since the color differences
being looked for in the CMD are at the level of a few hundredths of
a magnitude, precise definitions of the CMDs are necessary for this
method to give reliable results.
A principal conclusion from most of these studies was to show that
the most metal-poor clusters in the halo (those with [Fe/H] $\sim -2$)
have ages that were indistinguishable at the level of $\pm 1$ Gyr or less.
Although \cite{chab96} have argued on the basis of $\Delta V(MSTO-HB)$
that M68 is distinctly younger than M92, others such as \cite{car92a} and
\cite{vdb97} have used the same approach to suggest these two clusters
are nearly coeval.  Rather than relying on a single age-sensitive parameter
which may in practice be hard to define accurately for a given dataset,
we believe it is important to employ all relevant parts of the CMD
in an age comparison.  Our new data allow us
to define the fiducial sequences in the critical turnoff and subgiant
regions narrowly enough to perform similar tests on NGC 2419.

We can compare NGC 2419 {\it directly} only with clusters
that also have well defined CMD's in the $(V-I)$ plane, whereas
most previous photometry is in $(B-V)$ (cf. VBS90; SVB).
Fortunately, new data in $(V-I)$ are rapidly 
becoming available for many clusters including M92, which we will use as our
fiducial near-halo, low-metallicity cluster.  M92 has been the
point of comparison for several differential-age studies of
the low-metallicity clusters with excellent photometry, including
M15, NGC 4590, NGC 7099, and NGC 6397 (VBS90; SVB; Salaris \etal\ 1997).
The metallicity of M92, from a mean of
several recent measurements (\eg\ \cite{zin85}; \cite{bee90};
\cite{pet90}; \cite{sne91}; \cite{she96}), is [Fe/H] $=-2.25$. 
The uncertainty (precision) of the mean metallicity quoted in each of these studies is
typically $\pm0.06$ dex, but this value in most cases represents only the
internal precision of the particular method for measuring
line strengths.  A better estimate of the true external uncertainty 
(accuracy) of the metallicity can be obtained from the mutual agreement
among different studies, which for the five M92 analyses 
listed above is $\sigma$[Fe/H] $= \pm0.15$ dex (rms scatter).
Thus for M92 we estimate the uncertainty in the mean metallicity to be
near $\pm0.07$ dex.  By contrast, for NGC 2419 the metallicity measurement
relies principally on one study (\cite{sun88}), which gives [Fe/H] = $-2.14$ 
from low-dispersion measurements of Ca and Mg line strengths for eight stars. 
We adopt an uncertainty of $\pm0.15$ dex, assuming it to be comparable
to any of the single studies quoted above.  The difference between the two
clusters, $\Delta$[Fe/H](N2419-M92) = $0.11 \pm 0.17$, is small enough
that we can safely treat them as similar.

In Figure \ref{fig3}, we show the direct CMD
comparison between M92 and NGC 2419, using the new ground-based
$(V,V-I)$ photometry of M92 by \cite{bol97}.  The mean points for
M92 are listed in Table 2 (where the first 8 entries give the points
for the HB, and the remaining entries the giant branch and main sequence).  
In Fig.~\ref{fig3}, the mean points defining NGC 2419 
were shifted by the amounts $\Delta V, \Delta(V-I)$ shown 
in the figure so that the main sequences of the
two clusters coincide at a point 0.05 mag redder than the turnoff,
following the prescription of VBS90.  

By hypothesis, the horizontal shift $\Delta(V-I)$ represents the reddening
difference between the two clusters; thus, $E(V-I)_{N2419} = E(V-I)_{M92} + 0.14$.  
Similarly, the vertical shift
$\Delta V$ should represent their difference in distance moduli.  The value
$\Delta V = 5.28 \pm 0.04$ that we obtain by matching their main sequences
at the fiducial point just below the MSTO 
is, notably, quite similar to what we would have found by
matching only their horizontal-branch levels:  for M92, four high-quality 
photometric studies of the RR Lyraes and the HB give
$V(HB) = 15.15 \pm 0.03$ (\cite{car92b}; \cite{coh92b};
\cite{san69}; \cite{buo83}), whereas for NGC 2419, our new data give  
$V(HB) = 20.45 \pm 0.03$ from the HB stars nearest the RR Lyrae region.
The difference between the two is then $\Delta V(HB) = 5.30 \pm 0.04$, 
in close agreement with the offset employed in Fig.~3.  

The isochrones (described in more detail in \S4 below) used
for calibrating the differential ages are shown in Figure \ref{fig4}.
The color difference between the turnoff and lower giant branch,
for the metallicity of M92 or NGC~2419, changes 
at the rate $\Delta (V-I) /\Delta\tau = 0.013$ mag/Gyr over the 12- to 18-Gyr
age range shown; this ratio is quite insensitive to the 
chosen luminosity level on the giant branch, since the isochrone lines are
nearly parallel there.  The age sensitivity in the $(V-I)$ plane is 
similar to $(B-V)$, for which $\Delta (B-V) /\Delta\tau \simeq 0.012$
mag/Gyr at the same metallicity (VBS90).
From Fig. \ref{fig3}, we find that the observed
color difference between M92 and NGC~2419 is indistinguishable from zero 
at any level from the base of the RGB up to the HB $\sim 2$ mag higher.
We therefore adopt $\Delta(V-I) = 0.00 \pm 0.006$ mag, where the quoted
error is simply the precision in color with which we can perform the sliding fit
between the two colors.  In terms of age, this gives us
$\Delta \tau = 0.0 \pm 0.5$ Gyr.

A highly complementary way to use the differential CMD fit is to employ
the {\it magnitude} difference $\Delta V$ between the main-sequence
turnoff and the horizontal branch.  Since $\Delta V$ is a monotonically 
increasing function of cluster age, any age difference between the two
clusters in Fig. \ref{fig3} would be revealed as a vertical offset between the
HB levels, once the MSTO regions are superimposed.  That is, from the CMD fit
between the two clusters, we estimate the doubly differential quantity
$\Delta(\Delta V)$; if the fiducial sequences of the two clusters are as well
established as they are here, this quantity can readily be measured to within
a vertical uncertainty of $\pm 0.05$ mag.  It is apparent from
Fig. \ref{fig3} and also from the $\Delta V(HB)$ value calculated above that, 
to within this uncertainty, the offset $\Delta(\Delta V)$ 
between NGC 2419 and M92 is also zero.  The age sensitivity
of $\Delta V$ is $\simeq 0.073$ mag/Gyr for $\tau \sim 15$ Gyr (VBS96; 
see their Figure 4), which then translates 
into $\Delta \tau = 0.0 \pm 0.7$ Gyr, in close agreement with the color-shift
estimate.

In short, NGC 2419 {\it has essentially the same age as M92} if their
compositions are as similar as they appear to be.
By inference, NGC 2419 has the same age as
the other low-metallicity clusters in the 
Galactic halo to within the typical $\sim$1 Gyr precision of
the differential-age method.  These same objects  
are highly likely to be the oldest globular clusters (VBS90; VBS96; 
\cite{cha96}) as well as the oldest visible objects in the Galaxy
for which we can accurately measure ages.  The implication
is clear:  we are immediately forced to the conclusion that
{\it globular cluster formation began in the outermost halo of the
Galaxy at just as early a stage as it did in the inner parts of the halo}.

The only alternative we can suggest for this conclusion is that the
two clusters being compared here (NGC 2419 and M92) have 
{\it large, selective abundances differences} ([Fe/H], helium, or $\alpha-$elements)
which are working in conspiracy with an age difference  
to produce the identical CMD morphologies that we see.  
If the abundance difference is in metallicity, a 0.4-dex offset in [Fe/H]
would be required to mask a 1-Gyr age difference in the $(V-I)$ plane. 
Such a difference is too large to be accommodated
by the [Fe/H] measurements summarized above.
A 1-Gyr age decrease could also be produced by a helium abundance
increase of $\Delta Y \simeq 0.05$; again, existing data of several
kinds (see VBS96 for extensive discussion) make this option highly 
unlikely for an object this metal-poor.
The most plausible route to achieve an age difference may be in
the [$\alpha$/Fe] ratio, for which a 0.3-dex change in $\alpha$
would shift the deduced age by $\simeq 1$ Gyr.  If NGC 2419 is to be 
younger than M92, then it would
need to have [$\alpha$/Fe] $\gtsim 0.6$ if M92 has [$\alpha$/Fe] at the
`normal' level of +0.3.  Additional comments on this possibility
will be made below.

\section{Isochrone Fitting and Absolute Age Estimation}

Determining the {\it absolute} age of NGC 2419   
is then equivalent to asking:  What is the age of M92?
As will be seen below, this latter cluster provides the most incisive 
comparison between the observations and the theoretical stellar models 
because the foreground reddening, which is nearly negligible for M92,
is essentially eliminated as a free parameter for isochrone matching and
other comparisons with theoretical modelling.  

Given the well known uncertainty over RR Lyrae
luminosities and the Population II distance scale -- a controversy which
has been heightened by the recently published {\it Hipparcos} parallaxes 
for subdwarfs and Cepheids (\eg\, Reid 1997; \cite{fea97}) -- we will defer a 
full analysis of absolute ages.  In this paper, we will present only
sample isochrone fits which illustrate how 
the current stellar models match up
with the actual clusters under different assumed distance scales.
Since these stellar models, and their transformations into the observational
$(M_V, V-I)$ plane, will be used in our subsequent papers on the outer-halo
clusters, we first briefly describe their construction.

\subsection{Model Calculations}

     \cite{van97} have recently computed a large grid of
evolutionary sequences for low-mass, metal-poor stars that extend from the
Hayashi line through the main-sequence and red-giant phases to the zero-age
horizontal branch (ZAHB).  For each of the adopted [Fe/H] values between
$-2.3$ and $-0.3$, tracks were generated for [$\alpha$/Fe] $= 0.0, 0.3$, and
0.6, where the ``$\alpha$'' elements include O, Ne, Mg, Si, S, Ar, Ca, and
Ti.  Based on the \cite{zha90} and \cite{duf84} investigations, the 
abundances of Na and Cl were assumed to obey the relations
[Na/Fe] $=$ [Cl/Fe] $=$ [$\alpha$/Fe].
Aluminum and manganese were assumed to follow 
[Al/Fe] $=\,-\,$[$\alpha$/Fe] and
[Mn/Fe] $=\,-0.5\,$[$\alpha$/Fe] to approximate roughly the available data
(\eg\ \cite{whe89}; \cite{mag89}).
Finally, solar number abundance element-to-iron ratios were adopted for C, N,
Cr, and Ni (\cf\ \cite{whe89}), and the initial helium contents
were chosen to be consistent with $Y = 0.235 + 1.936\,Z$;  
for the [Fe/H] values of
concern here (NGC 2419 and M92), this assumption produces a negligible
increase in $Y$ over the cosmological value.
 
    Opacities similar to those reported by \cite{rog92} for
temperatures $\ge 6000$~K and to those given by \cite{ale94}
for lower temperatures were computed for the adopted element mixes (see
VandenBerg \etal\ 1997 for details).  The relatively minor improvements
to the H-burning nuclear reaction rates favored by Bahcall \& Pinsonneault (1992)
and a treatment of Coulomb interactions in the equation of state
represent the only substantive changes to the stellar evolution code,
compared with that described by VandenBerg (1992, and references
therein).  All calculations assumed a value of $\alpha_{\rm MLT} = 1.89$ for
the usual mixing-length parameter, to be consistent with the requirements of
a Standard Solar Model, and the surface pressures were derived by integration of
the hydrostatic equation in conjunction with the \cite{kri66} $T$-$\tau$
relation.  Using model atmospheres to derive the boundary pressures would
clearly have been the preferred approach, but such calculations for the
required wide range in gravity, effective temperature, [Fe/H], and
[$\alpha$/Fe] are not yet available.  In spite of this deficiency, \cite{van97}
demonstrate that the predicted $T_{\rm eff}$ scale of their
models appears to agree quite well with existing observational constraints.
Isochrones on the theoretical plane were obtained by interpolation in these
tracks with the methods described by \cite{ber92}.  

\subsection{Model Transformations to the Observational Plane}

The approach we adopt to convert the theoretical isochrones ($M_{bol}, T_{eff}$)
to the observational plane ($M_V, V-I$) is slightly different from that used in
other recent discussions, such as VBS96, where the transformation
to observed quantitities was accomplished with semi-empirical
bolometric corrections and color-$T_{eff}$ relations.
Here, we take the more classical route of 
{\it transforming the isochrones as strictly as possible through 
stellar-atmosphere models}; \ie, we carry both the observations and
the theory as far as they can go on their own ground, and only then
do we compare them directly.  

   Values of log $g$ were found at 50 K increments in $T_{eff}$
for $T_{eff} > 5000$ K, at 100 K increments for $T_{eff} < 5000$ K, and at the
turnoff (hottest) temperature, for each of the sets of isochrones
through cubic spline interpolation.  The opacity distribution functions
(ODFs) were found for the isochrone abundances by interpolation in ODFs
with abundances of $-3.0$, $-2.0$, $-1.0$ and $-0.5$.  The Marcs program
(\cite{gus75}) was then used to calculate model atmospheres
for each of these ($T_{eff}$, log $g$) points.  
The H/He/metals ratios used in
the model atmosphere and synthetic spectrum calculations were the same as
those employed in the stellar interior work.  The models calculated for
enhanced alpha-element abundances, i.e. [$\alpha$/Fe] = +0.6 and +0.3,
used the same ODFs as those with [$\alpha$/Fe] = 0.0.  
However, the $P_g - P_e - T$ relationships were calculated 
in the models allowing for the different [$\alpha$/Fe] values.

    The model atmospheres were then used for synthetic spectrum
calculations, again allowing for the different $\alpha$-element
abundances.  The microturbulent contribution to the Doppler broadening
velocity was chosen to vary with log $g$ by interpolating in the values
1.0 km/sec at log $g = 4.5$; 1.7 km/sec at log $g = 1.5$; 
and 2.5 km/sec at log $g = 0.5$.  The line list was an 
improved version of that used by \cite{bel94} and \cite{tri95}, who
give examples of fits to the spectra of the Sun and Arcturus.

    The synthetic spectra were multiplied by the pass band sensitivity
functions and were converted to magnitudes to give the surface brightness
magnitudes of the models. The sensitivity functions 
were the \cite{bes90} ones
for $UBVRI$ and the WFPC2 filters F555W and F814W.  The magnitude zero points
were found by requiring the magnitudes of the \cite{dre81} Vega
model to match the fluxes given by \cite{hay85}, then finally by matching them
to Bessell's (1983) ($V, V-I$) observational data for Vega.
Normalizing the Dreiling-Bell model to the Hayes fluxes required a 
$-0.004$ magnitude adjustment to the model $(V-I)$ colors.
The Bessell data have also been used for WFPC2 zero points by
\cite{hol95}.

	The Bessell $V$ magnitude results are very similar to those
calculated for the F555W pass band:  the difference in 
absolute visual magnitude for virtually all
the models is $< 0.01$ mag.  This is in agreement with earlier WF/PC
calculations for these pass bands by \cite{edv89}.  The $I$
band and the F814 magnitudes also agree very well, again following the
results of Edvardsson \& Bell.  This agreement
is somewhat surprising, in view of the difference in sensitivity 
function profiles, in the detectors, and in the effect of $H_2O$ and 
$O_2$ telluric lines on ground-based data in this spectral region.

    The isochrones are based upon a solar apparent visual magnitude of
$V = -26.73$, which gives $M_V = 4.84$, and an adopted solar bolometric
correction of $-0.12$.  The $T_{eff}$ and 
visual surface brightness of a solar
model and this bolometric correction were used to find the bolometric
corrections of the cluster models from their $T_{eff}$ and visual surface
brightnesses.  
The zero point of the bolometric correction scale consequently rests on the 
visual surface brightness of this solar model. We note that \cite{bel96}
use the angular diameters of the Sun and Vega, the visual
surface brightnesses of their models and the apparent magnitude of Vega  to
derive $V = -26.77$ for the Sun. This value is consistent with that derived
from solar fluxes and from direct observation of the Sun ($V = -26.75 \pm 0.06$;
\cf\ Hayes 1985). However, uncertainties in the angular diameter of Vega, the 
solar model and solar observations must translate into an uncertainty of 
perhaps $\pm 0.05$ mag in the bolometric corrections of field subdwarfs
and consequently into comparisons of isochrones and cluster main
sequences.  The visual surface brightness magnitudes of the metal-poor
giant models are brighter than those of Population I models of the
same $T_{eff}$.  This brightening is due in part 
to the smaller line blocking and in part to the differences in 
model structure, which cause the continuous flux
of the metal-poor models to be greater.  This effect 
in turn causes the bolometric
corrections for the metal-poor models to be smaller than those of the
Population I models.  These changes also cause the $(V-I)$
colors of the metal-poorer models to be slightly redder than 
those of the metal-richer models of the same $T_{eff}$.

    The properties of the solar model ($T_{eff}= 5760$ K, 
log $g =4.44$) that was used to set the zero 
point of the bolometric correction are discussed by \cite{bel96}.
This model is fainter in $V$ than that used by VandenBerg \& Bell (1985, hereafter
VB85), owing to the addition of further line and continuous opacity sources which are
dependent on metal abundance.  These opacity sources have less effect at
lower metallicity and so the $V$ magnitudes of the present metal-poor 
models agree more closely with the VB85 ones than do the  solar models. 
Consequently the bolometric corrections of the metal-poor models are
$\sim 0.1$ mag larger in an absolute sense than those of 
VB85, causing a decrease of $\sim 1.5$ Gyr at a fixed turnoff luminosity.
Further discussion of this point can be found in \cite{vdb97}.
 
	The $(V-I)$ colors of the giant branch models are slightly bluer
than those of the dwarfs at the same $T_{eff}$, 
by an amount which increases with decreasing $T_{eff}$.  
The cooler $\alpha = +0.6$ models have somewhat bluer colors than
those with $\alpha = 0.0$.  The $(V-I)$ colors for the [Fe/H] = $-2.14$
isochrones are very similar to those published by VB85
for dwarfs for [Fe/H] = $-2.0$ and by \cite{bel89} for
giants of the same abundance. 
The surface abundances of some metals, particularly C (\cite{bel79}),
alter as stars evolve along the giant branches of some
metal-poor globulars. However, we have not allowed for this, since the effect 
on broadband colors is expected to be small.

\subsection{The Age of M92 and NGC 2419:  Old or Young?}

The transformed isochrones can now be superimposed on the CMD for each
cluster.  The models shown here have 
a composition [Fe/H] = $-2.14$, $Y = 0.235$, and [$\alpha$/H] = 0.3.
The moderate $\alpha-$enhancement is supported by direct spectroscopic
measurement of oxygen abundances in M92 and the great majority of other metal-poor clusters
(Carney 1996; \cite{sne91}), though in a few other objects (notably M13;
see \cite{pil96}; \cite{kra97}) an [$\alpha$/Fe] ratio closer to the solar value
is observed.

In Figure \ref{fig5}, we show the match between the transformed
isochrones and the M92 CMD, where the distance scale is set essentially
by fitting the model ZAHB at the level of the M92 horizontal branch.
The cluster CMD is then shifted horizontally until the unevolved main
sequence matches the models.  If the model colors are correct,
the color shift subtracted from the cluster mean points should then represent the 
reddening.  Our deduced shift of $\delta(V-I) = 0.013 \pm 0.01$ (estimated
uncertainty of fit) corresponds to $E(B-V) = 0.01 \pm 0.01$, which is in
close agreement with the normally used value for M92 of $E(B-V) = 0.02$ (\cite{san69};
\cite{har96}) and is consistent with the idea that the model colors do not
need further arbitrary zero-point adjustment.

The resulting distance modulus, $(m-M)_V = 14.60 \pm 0.06$,
corresponds to a horizontal-branch luminosity at the level of
the RR Lyraes of $M_V(HB) = 0.55$.  This level agrees to within
$0.1$ magnitude with:  (a) previous calibrations from subdwarf parallaxes
(\eg\ VBS96; \cite{san96}); (b) ZAHB models (\cite{lee90}; \cite{dor92});
(c) Cepheid distances to the RR Lyrae populations in Local Group dwarf
galaxies including the LMC, SMC, and IC 1613 (\cite{van95b}; \cite{arw92};
\cite{sah92}); and (d) the distance modulus to the moderately metal-poor
cluster NGC 6752 (at [Fe/H] $= -1.6$) calibrated through its white-dwarf 
sequence and nearby field white dwarfs (\cite{ren96}).  
It is, however, $0.1 - 0.2$ magnitudes brighter than
HB luminosities measured from (a) the Baade-Wesselink method (\eg\ Carney \etal\ 1992a);
(b) Cepheid-calibrated distances to the RR Lyraes in M31 (\cite{fus96}); and
(c) statistical parallax of field RR Lyraes (\cite{lay96}).\footnote{We note
that the Layden \etal\ mean
value of $M_V = 0.71 \pm 0.12$ at [Fe/H] $=-1.6$ would rise to $M_V \sim 0.61$
if the \cite{stu66} prescription for calculating the RR Lyrae
reddenings were adopted.  Normalizing this value to [Fe/H] $= -2.2$
using $\partial M_V/\partial{\rm [Fe/H]} = 0.15$ (Carney \etal\ 1992a),
we would obtain $M_V \simeq 0.52$ at the metallicity of M92, in good agreement
with the present estimate.}  Recent {\it Hipparcos} parallaxes of a few 
blue-HB field stars (\cite{deb97}) give $M_V(HB) \sim 0.7 \pm 0.2$, a value
which is consistent with any of the other methods listed above.

In Figure \ref{fig6}, the same isochrone fit is shown for
NGC 2419.  The resulting reddening estimate (again, under the assumption
that the model colors along the main sequence are systematically accurate)
is $E(V-I) = 0.145 \pm 0.01$ and thus $E(B-V) = 0.11 \pm 0.01$.  To within
its quoted uncertainty, 
this estimate agrees with the differential color shift from Fig.~3 added
to the (small) reddening of M92.  It is also consistent with the value of
$E(B-V) = 0.10 \pm 0.05$ obtained by \cite{chr88} by CMD fitting to both 
M15 and M92 in the $(V,B-V)$ plane. The distance modulus of $(m-M)_V = 19.88 \pm 0.06$
gives $M_V(HB) = 0.57$, and a true distance for NGC 2419 of 
81 kpc from the Sun or $\simeq 90$ kpc from the Galactic center.

For both clusters, the best-fitting age read off the isochrones is
$(15 \pm 1)$ Gyr.  VBS96 and \cite{vdb97} obtained similar results
from isochrone fitting in the $(B-V)$ plane, by using a semi-empirical
color transformation procedure somewhat different from the model-atmosphere
transformations that we employ here.  A further reduction of $\ltsim 1$ Gyr 
might be obtained by incorporating a realistic amount of helium diffusion
into the stellar models, as discussed by \cite{pro91}, \cite{vdb97},
and \cite{cas97}.
More drastic changes than this now seem very hard to achieve in a natural
way within the context of the most recent models; see VBS96 and \cite{vdb97}
for more detailed discussion of the input physics.  

The overall quality of fit of the isochrones to the cluster data, in both
Figs.~5 and 6, is virtually identical with what we would have obtained 
under various other assumptions.  For example, instead of setting the 
distance by matching the ZAHB to the cluster HB, we could have subtracted the 
(observationally known) reddening from the
cluster mean points and then shifted it vertically until the {\it main sequence} fell
in line with the models.  Within the quoted uncertainties, the same answers emerge
for the distance modulus and reddening.  Still another approach that relies
even more heavily on the correctness of the models would be to
{\it perform an unconstrained fit of the CMD to the models independently of
other observational input}:  that is, we could find the distance
modulus and reddening that give the best `global' match of all features in
the CMD to the isochrones (main sequence, subgiants, horizontal branch, 
and giant branch).  The results are again the same as before to within the
internal uncertainties of the method.

The concordance between the model isochrones and the real clusters
continue to improve with advances in both the theory and the data,
but residual discrepancies show up in three areas (Figs.~5 and 6):
(a) the ZAHB model line should, ideally,
lie $\sim 0.05 - 0.1$ mag fainter than the mean HB points at the RR Lyrae
region to take account of post-ZAHB luminosity evolution (\eg\ \cite{lee90};
\cite{dor92}; Salaris \etal\ 1997); (b) the theoretical RGB 
line runs nicely parallel to the observed giant branch but 
is consistently redder by $\Delta(V-I) \simeq 0.02 - 0.03$ mag;
and (c) along the turnoff and subgiant region, the mean data points cross over
two isochrone lines, starting approximately on the 16-Gyr line at the MSTO
and finishing on the 14-Gyr line at the base of the giant branch.
All of these discrepancies are at such a low level that
we speculate that very small
uncertainties in the photometric zeropoints, the reddening, the abundances,
the isochrones themselves, and
the model transformations (at the level of $0.01 - 0.02$ mag in each)
may have conspired to leave the various offsets that we see.

Many other choices of parameters -- distance modulus, reddening, 
$\alpha-$abundance, etc. -- which differ only slightly from 
the ones shown above can be made, with quite plausible results. 
An exhaustive exploration of this parameter space will not
be presented here, but as an illustration of the possibilities, 
we show another sample isochrone fit in Figure \ref{fig7} for NGC 2419:
here, the reddening and distance modulus have been deliberately
chosen to produce the theoretically expected `ideal' match strictly for both
the horizontal branch (where the ZAHB is placed $\simeq 0.07$ mag fainter than
the observed mean HB) and the unevolved main sequence (where the isochrone
line runs exactly through the mean main-sequence points for the entire range $M_V > 4$).
The best-fitting age is now 14 Gyr, and the remaining discrepancy in the fit has now
been put entirely on the theoretical RGB, which stands redward of the real cluster
by $\simeq 0.05$ mag.  This solution requires a cluster reddening of
$E(B-V) = 0.12$ and an RR Lyrae luminosity of $M_V(HB) = 0.50$.  Overall, reasonable
fits to the data can be found for isochrone shifts that differ from one solution
to the next by $\pm 0.02$ mag in color and $\pm 0.1$ mag in luminosity.
The corresponding (external) uncertainty in the cluster age is then $\pm 2$ Gyr.

\cite{sal97} provide another recent analysis of
the ages of the most metal-poor clusters
(M15, M68, M92), using isochrones from their independently calculated set 
of stellar models.  By using the ZAHB to set the distance moduli, 
they employ only the deduced MSTO luminosity to estimate the cluster ages, 
and find all three clusters to
lie in the range $(12 \pm 2)$ Gyr.  Aside from details of the 
codes and opacity prescriptions, the principal
differences between their models and those of \cite{van97} appear to be in
the adopted abundances (Salaris \etal\ use a value 
[$\alpha$/Fe] = +0.5 which is on the upper end of the plausible range) 
and in the luminosities of the ZAHB models (the Salaris \etal\ HB models are brighter
by $\sim 0.10 - 0.15$ mag).  
These effects generate most of the $\sim 2$ Gyr difference in ages that we find
for the same clusters.  Neither set of models includes diffusion (Proffitt \& 
VandenBerg 1991; Castellani \etal\ 1997).  Nevertheless, the comparison
suggests that the true internal uncertainties in the models, given
identical input parameters, are at the $\pm 1$ Gyr level (see Gratton \etal\
1997 for a similar conclusion).

However, changes to the distance scale have recently been proposed that go
well beyond the range discussed above.  \cite{rei97} has used new parallaxes
of 15 low-metallicity subdwarfs from
the {\it Hipparcos} database to fit the main sequences of five metal-poor
globular clusters, thus calibrating their HB luminosities.
For M5 and M13 (at [Fe/H] $\simeq -1.5$), he obtains $M_V(HB) \simeq 0.55$,
quite similar to the levels obtained from the 
unconstrained isochrone fits that we discussed above.
However, for M92, M15, and M68 (at [Fe/H] $\simeq -2.1$), he obtains
$M_V(HB) = 0.15$, about 0.4 mag brighter.
In Figures \ref{fig8} and \ref{fig9}, we show the implications of 
a ZAHB luminosity this high.  The best-fitting age would now      
be near 10 Gyr, extrapolating from the four isochrones plotted
(or $\sim 9$ Gyr after accounting for helium diffusion).
But the overall isochrone fit is obviously
seriously discrepant in three ways:  (a) This solution would require the
models to have an arbitrary color adjustment of $\delta(V-I) \simeq 0.07$
mag {\it on the main sequence} even
after the true cluster reddening is accounted for, in the sense that the predicted
model colors are too blue by that amount.  (b) The ZAHB model line is too faint
by $\sim 0.4$ mag.  (c) The RGB line stands off the cluster points by almost
0.1 mag.  Curiously, this means that the model colors for the {\it giants}
would be nearly correct as they stand, and that the {\it main-sequence} model
colors would be the ones that require a large arbitrary redward correction.
Normally, the model RGB colors are taken as the more easily adjustable because
of their strong dependence on the modelling of convection.

\cite{gra97} have also used {\it Hipparcos} parallax data for 7
low-metallicity subdwarfs to calibrate the distances and ages of
several clusters; for the lowest-metallicity clusters, they find
$M_V(HB) \simeq 0.2 - 0.3$, a level about halfway between our estimates
and those of \cite{rei97}.  Again using only the MSTO luminosity to
calibrate the age, they derive $t \sim (14 \pm 1)$ Gyr for the oldest clusters
in their sample (M92, NGC 288, and NGC 6752, with the results somewhat
dependent on which of several conversion models is adopted; see their Table 2).

These high estimated HB luminosities for M92 and the other low-metallicity
clusters may turn out to be an artifact of the small number of stars in
subdwarf samples, along with a variety of other biases such as the
presence of binaries and the [Fe/H] measurements themselves.  
These issues are discussed in
a more recent analysis of the {\it Hipparcos} subdwarf parallaxes 
by \cite{pon97}, based on a considerably larger sample of metal-poor stars 
and a more extensive analysis of biases including radial velocity data to
detect binaries.  \cite{pon97} derive $(m-M)_V = 14.67$ and thus $M_V(HB) = 0.48$
for M92, a value quite similar to what we find purely from the isochrone fits
(Fig.~\ref{fig5}).  In summary, the full impact of the new {\it Hipparcos} data,
and the continuing improvements to the stellar models, has yet to be felt;
nevertheless, we believe that an age in the generous range of 12 to 15 Gyr 
for the most metal-poor clusters in the Galaxy is well supported by the
current mix of theory and observation.

\section{Discussion and Summary}

Our WFPC2 photometry for NGC 2419 allows us to define the
CMD loci for this remote-halo, 
low-metallicity cluster to a level three magnitudes below the 
main-sequence turnoff.  Our analysis of the CMD
shows that it has the same age, to within $\sim 1$ Gyr,
as M92 and other mid-halo clusters which have similarly low  metallicity
and color-magnitude morphology.
For this low-[Fe/H] subgroup of clusters at least, we therefore find no
detectable age gradient through the Galactic halo from $R_{gc} \simeq
7$ kpc to 90 kpc (\cf\ Richer \etal\ 1996 and \cite{wei97} 
for additional discussion).  
The clear implication is that all parts of the Milky Way protogalaxy {\it began}
their earliest star formation at very much the same time.

The differential-age determinations that we have
employed here may be vitiated by {\it large} differences in
composition between NGC 2419 and M92, most notably in the $\alpha-$element ratios.
No evidence for such differences was found by Suntzeff \etal\ (1988) from
7.4 \AA-resolution blue spectra of nine giants in NGC 2419; nor did
\cite{car96} find large differences between clusters, 
from his review of the bulk of the available 
evidence for halo and globular cluster stars.
Nonetheless, until high-dispersion spectroscopic abundance analyses
can be carried out for NGC 2419, abundance differences in the
$\alpha-$elements cannot be ruled out definitively.
If NGC 2419 is indeed younger than the inner-halo objects of similar
metallicity, quite a high [$\alpha$/Fe] level ($\gtsim 0.6$) will be
required.

In addition, we have still not obtained high-quality
age determinations for any low-metallicity clusters in the 
{\it innermost} $\sim5$ kpc of the Galaxy, which is the one
remaining region where significantly older clusters
might still lurk undetected.  But in this respect it should also be noted that
there are only a handful of clusters known with [Fe/H] $\ltsim -1.7$ 
{\it and} $R_{gc} < 5$ kpc (see, \eg\, \cite{ste94} and the more
recent data from the catalog of Harris 1996).  In addition, {\it all} of these few 
have high radial velocities which rule out the possibility
that they spend most of their time within the bulge.  These bits of evidence,
though not definitive, suggest
that there are very few extremely low-metallicity clusters that genuinely 
belong to this innermost region.  Nevertheless, it is not
out of the question that older objects could exist even if their metallicities
are not extremely low.  Either way, the age distribution of the 
inner-halo clusters needs to be explored more fully.

We have also shown sample fits of the CMDs for M92 and NGC 2419 to
up-to-date isochrones incorporating well calibrated transformations
to the observational $(M_V,V-I)$ plane.  
Using the distance scale that most naturally fits the ZAHB model
luminosity {\it and} the main-sequence colors, we find that the best-fitting 
age for the most metal-poor globular clusters in the Milky Way is near 14 Gyr.

Our picture of the earliest epoch of the Galaxy is one 
in which clusters began to appear at very much the same time
everywhere across a vast protogalactic region, spanning perhaps 
200 kpc diameter in present-day dimensions.  
We do {\it not}, however, necessarily conclude that the 
near-simultaneous formation of all these widely spread `first' clusters 
was therefore coordinated globally by some ELS-style
monolithic collapse.  Considerable evidence based on both the metallicities 
(Searle \& Zinn 1978) and masses (\cite{har94}) of the halo globular 
clusters supports the view that they were born as protoclusters embedded within host
$\sim 10^8 - 10^9 M_{\odot}$ gas clouds (the ``supergiant molecular
clouds'' or SGMCs of \cite{har94})
and not just as isolated condensations within the greater proto-Galactic
halo.  If clusters form within these primeval dwarf-galaxy-like
gaseous fragments, then (see \cite{har94}; \cite{mcl96})
their formation timescale is determined by the time needed to build up
$10^5 - 10^6 M_{\odot}$ dense gas clouds within these SGMCs,  
and not (as in ELS) the free-fall time of the whole protoGalactic region.
This timescale for protocluster growth is typically 
a few $10^8$ y or less, depending on the density and external pressure
of the SGMC, but even in the larger, lower-density
clouds the growth time is typically $\ltsim 1$ Gyr.

In this picture, the observation that the most metal-poor 
globular clusters have essentially the same age everywhere in the halo 
has a different interpretation from the classic ELS 
scenario, but an equally simple one:  it requires that
all of the various primordial SGMCs that would eventually 
merge to build the larger Galaxy must have {\it begun} building 
the first generation of stars and clusters in the same $\sim 1-$Gyr
time period.  We suggest that this requirement can be automatically
satisfied by cosmological boundary conditions.  The first
discrete gaseous structures would have emerged everywhere
at redshifts $z \gtsim 5$, and current models (\eg, \cite{kau93};
\cite{sil93})
predict that the mass spectrum of the emergent clouds should 
peak in the dwarf-size region ($10^8 - 10^9 M_{\odot}$).
In other words, these clouds would have been set in place,
scattered throughout the potential wells of larger protogalaxies,
within about a Gyr of the recombination epoch.
Inside the SGMCs, formation of massive, protoglobular 
clusters could then have immediately begun no matter where they 
found themselves, yielding fully formed clusters within a few 
$10^8$ years later, Protocluster buildup would necessarily have proceeded 
faster in the higher-density, higher-pressure SGMCs in the centermost regions
of large protogalaxies, but within a $\pm 1-$Gyr age spread the 
differences would now be indistinguishable.

These initial conditions would, in short, allow globular cluster formation to  
begin in all parts of the larger protogalaxy at very
much the same time and, {\it even without
a rapid global collapse of the protohalo,} we might expect the first generation
of clusters to have formed within the narrow age spread that
we now observe.  As is noted by Harris \& Pudritz, this same model 
of cluster formation within SGMCs (whose sizes are constrained by external
pressure) also predicts the systematic increase of cluster half-mass radius
with Galactocentric distance in just the proportions that we see
($r_h \sim R_{gc}^{1/2}$).

If such a description of these early events were true, we could scarcely claim
that the Milky Way is in any way unique. We would
therefore expect that the oldest globular clusters in any large galaxy,
spiral or elliptical alike, should possess very much the same age
as we have found for the Milky Way.

\acknowledgments

The research of WEH, GGF, HBR, and DAV is supported through grants from
the Natural Sciences and Engineering Research Council of Canada.
Support was provided to RAB and HEB by NASA through grant number
GO-05481.02-93A from the Space Telescope Science Institute, which is
operated by the Association of Universities for Research In Astronomy, Inc.,
under NASA contract NAS5-26555.  RAB also acknowledges support from 
NSF grant AST93-14931.

\clearpage
\begin{deluxetable}{rrrrr}
\tablenum{1}
\tablecaption{Fiducial Points for NGC 2419 CMD \label{tab1}}
\tablewidth{0pt}
\tablecolumns{5}
\tablehead{
\colhead{$V$} & \colhead{$\pm$} & \colhead{$V-I$} &
\colhead{$\pm$} & \colhead{n}
}
\startdata
  18.816 &     .043 &    1.111 &     .005 &   10 \nl
  19.228 &     .027 &    1.070 &     .005 &   17 \nl
  19.614 &     .028 &    1.016 &     .006 &   19 \nl
  19.969 &     .031 &     .932 &     .015 &   11 \nl
  20.337 &     .057 &     .916 &     .009 &    7 \nl
  20.431 &     .090 &     .783 &     .005 &    5 \nl
  20.381 &     .049 &     .757 &     .007 &    3 \nl
  20.378 &     .052 &     .705 &     .004 &    3 \nl
  20.507 &     .049 &     .393 &     .007 &    4 \nl
  20.618 &     .048 &     .224 &     .006 &    4 \nl
  20.645 &     .016 &     .192 &     .003 &   12 \nl
  20.687 &     .013 &     .151 &     .002 &   38 \nl
  20.788 &     .019 &     .114 &     .003 &   16 \nl
  20.922 &     .030 &     .073 &     .004 &   16 \nl
  20.993 &     .067 &     .034 &     .004 &    3 \nl
  21.369 &     .026 &    -.002 &     .009 &   11 \nl
  21.713 &     .039 &    -.025 &     .010 &   11 \nl
  22.572 &     .014 &    -.034 &     .010 &   12 \nl
  24.211 &     .074 &    -.236 &     .024 &   10 \nl
  24.938 &     .028 &    -.211 &     .042 &   16 \nl
 \nl
  17.499 &     .023 &    1.372 &     .016 &    7 \nl
  17.710 &     .016 &    1.304 &     .012 &    7 \nl
  17.918 &     .025 &    1.288 &     .013 &    8 \nl
  18.085 &     .018 &    1.255 &     .009 &   10 \nl
  18.293 &     .023 &    1.212 &     .004 &    9 \nl
  18.516 &     .019 &    1.211 &     .007 &   12 \nl
  18.719 &     .015 &    1.194 &     .005 &   16 \nl
  18.891 &     .012 &    1.166 &     .006 &   19 \nl
  19.103 &     .013 &    1.161 &     .008 &   21 \nl
  19.310 &     .010 &    1.132 &     .005 &   25 \nl
  19.504 &     .011 &    1.106 &     .004 &   36 \nl
  19.726 &     .009 &    1.094 &     .003 &   37 \nl
  19.905 &     .008 &    1.088 &     .003 &   59 \nl
  20.102 &     .008 &    1.073 &     .006 &   41 \nl
  20.309 &     .008 &    1.049 &     .002 &   60 \nl
  20.489 &     .007 &    1.042 &     .003 &   61 \nl
  20.705 &     .007 &    1.026 &     .003 &   74 \nl
  20.908 &     .007 &    1.021 &     .003 &   81 \nl
  21.066 &     .013 &     .974 &     .017 &    7 \nl
  21.146 &     .011 &     .982 &     .023 &    9 \nl
  21.248 &     .009 &     .991 &     .006 &   10 \nl
  21.371 &     .008 &     .998 &     .013 &    8 \nl
  21.455 &     .010 &     .979 &     .008 &   15 \nl
  21.549 &     .012 &     .971 &     .011 &    8 \nl
  21.661 &     .007 &     .965 &     .010 &   12 \nl
  21.749 &     .009 &     .973 &     .008 &   12 \nl
  21.865 &     .007 &     .980 &     .005 &   16 \nl
  21.951 &     .006 &     .959 &     .007 &   21 \nl
  22.041 &     .008 &     .937 &     .017 &   14 \nl
  22.142 &     .007 &     .946 &     .008 &   16 \nl
  22.239 &     .006 &     .943 &     .007 &   18 \nl
  22.357 &     .007 &     .936 &     .006 &   19 \nl
  22.448 &     .006 &     .936 &     .007 &   27 \nl
  22.559 &     .007 &     .917 &     .010 &   20 \nl
  22.647 &     .005 &     .928 &     .007 &   29 \nl
  22.759 &     .004 &     .905 &     .006 &   31 \nl
  22.846 &     .005 &     .916 &     .005 &   29 \nl
  22.949 &     .005 &     .885 &     .008 &   37 \nl
  23.045 &     .005 &     .866 &     .007 &   36 \nl
  23.149 &     .004 &     .842 &     .004 &   58 \nl
  23.248 &     .003 &     .798 &     .005 &   71 \nl
  23.352 &     .004 &     .754 &     .005 &   78 \nl
  23.447 &     .003 &     .729 &     .004 &   77 \nl
  23.550 &     .003 &     .705 &     .004 &  101 \nl
  23.651 &     .003 &     .694 &     .003 &  119 \nl
  23.750 &     .003 &     .688 &     .003 &  127 \nl
  23.853 &     .002 &     .684 &     .003 &  174 \nl
  23.951 &     .002 &     .678 &     .002 &  185 \nl
  24.051 &     .002 &     .677 &     .003 &  208 \nl
  24.148 &     .002 &     .691 &     .003 &  211 \nl
  24.252 &     .002 &     .692 &     .003 &  209 \nl
  24.350 &     .002 &     .694 &     .003 &  242 \nl
  24.452 &     .002 &     .701 &     .003 &  235 \nl
  24.553 &     .002 &     .715 &     .003 &  264 \nl
  24.650 &     .002 &     .722 &     .003 &  295 \nl
  24.754 &     .002 &     .730 &     .003 &  307 \nl
  24.850 &     .002 &     .736 &     .004 &  305 \nl
  24.949 &     .001 &     .747 &     .003 &  356 \nl
  25.050 &     .001 &     .764 &     .004 &  322 \nl
  25.152 &     .001 &     .774 &     .004 &  362 \nl
  25.250 &     .001 &     .781 &     .004 &  406 \nl
  25.349 &     .002 &     .802 &     .004 &  409 \nl
  25.451 &     .001 &     .803 &     .004 &  400 \nl
  25.551 &     .002 &     .828 &     .004 &  480 \nl
  25.650 &     .001 &     .838 &     .004 &  443 \nl
  25.748 &     .001 &     .854 &     .004 &  466 \nl
  25.850 &     .003 &     .871 &     .005 &  460 \nl
  25.949 &     .001 &     .889 &     .005 &  522 \nl
  26.050 &     .001 &     .908 &     .005 &  466 \nl
  26.150 &     .002 &     .929 &     .006 &  465 \nl
  26.251 &     .001 &     .935 &     .006 &  492 \nl
  26.348 &     .001 &     .972 &     .006 &  507 \nl
  26.451 &     .001 &     .992 &     .007 &  477 \nl
  26.549 &     .001 &    1.000 &     .007 &  478 \nl
  26.653 &     .001 &    1.010 &     .007 &  455 \nl
  26.751 &     .001 &    1.061 &     .008 &  481 \nl
  26.850 &     .002 &    1.063 &     .009 &  444 \nl
  26.950 &     .001 &    1.111 &     .009 &  412 \nl
  27.049 &     .003 &    1.077 &     .012 &  370 \nl
  27.151 &     .001 &    1.118 &     .013 &  352 \nl
  27.250 &     .002 &    1.115 &     .013 &  329 \nl
  27.349 &     .002 &    1.192 &     .014 &  252 \nl
  27.451 &     .001 &    1.156 &     .015 &  241 \nl
  27.550 &     .002 &    1.202 &     .015 &  209 \nl
  27.647 &     .002 &    1.206 &     .016 &  157 \nl
  27.748 &     .003 &    1.243 &     .023 &   99 \nl
\enddata
\end{deluxetable}

\clearpage
\begin{deluxetable}{rrrrrr}
\tablenum{2}
\tablecaption{Fiducial Points for M92 CMD \label{tab2}}
\tablewidth{0pt}
\tablecolumns{4}
\tablehead{
\colhead{$V$} & \colhead{$V-I$} & \colhead{} &
\colhead{$V$} & \colhead{$V-I$}
}
\startdata
   16.62 & -0.120 && 17.81 &  0.735  \nl
   16.12 & -0.084 && 17.90 &  0.689  \nl
   15.83 & -0.052 && 17.96 &  0.649  \nl
   15.57 & -0.019 && 18.09 &  0.613  \nl
   15.40 &  0.017 && 18.17 &  0.592  \nl
   15.33 &  0.070 && 18.26 &  0.576  \nl
   15.24 &  0.143 && 18.39 &  0.561  \nl
   15.17 &  0.228 && 18.54 &  0.557  \nl
         &        && 18.69 &  0.555  \nl
   12.75 &  1.143 && 18.89 &  0.558  \nl
   13.90 &  1.022 && 19.06 &  0.568  \nl
   14.67 &  0.972 && 19.37 &  0.589  \nl
   15.91 &  0.867 && 19.67 &  0.617  \nl
   16.61 &  0.828 && 19.95 &  0.648  \nl
   17.09 &  0.810 && 20.65 &  0.740  \nl
   17.39 &  0.781 && 21.11 &  0.816  \nl
   17.60 &  0.766 && 21.58 &  0.913  \nl
   17.74 &  0.754 && 22.13 &  1.036  \nl
\enddata
\end{deluxetable}

\clearpage
\begin{center}
Figure Captions
\end{center}

\figcaption[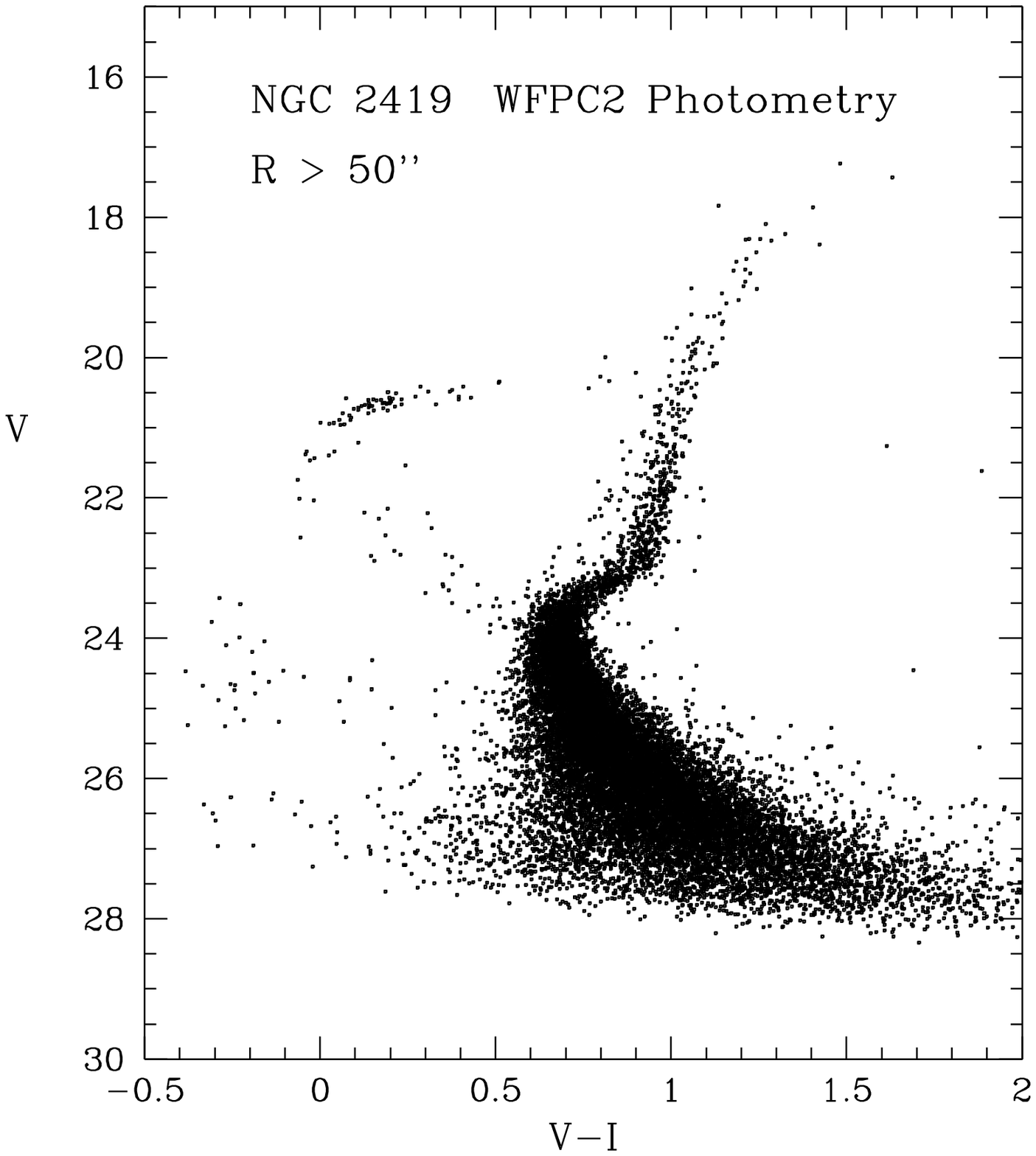]{Composite color-magnitude diagram for NGC 2419,
from HST/WFPC2 images and ALLFRAME data reduction; see \protect\cite{ste97}.
Stars further than $50''$ from cluster center are plotted, selected as
described in \S2 of the text.
This CMD is employed for definition of the fiducial sequences.  \label{fig1}}

\figcaption[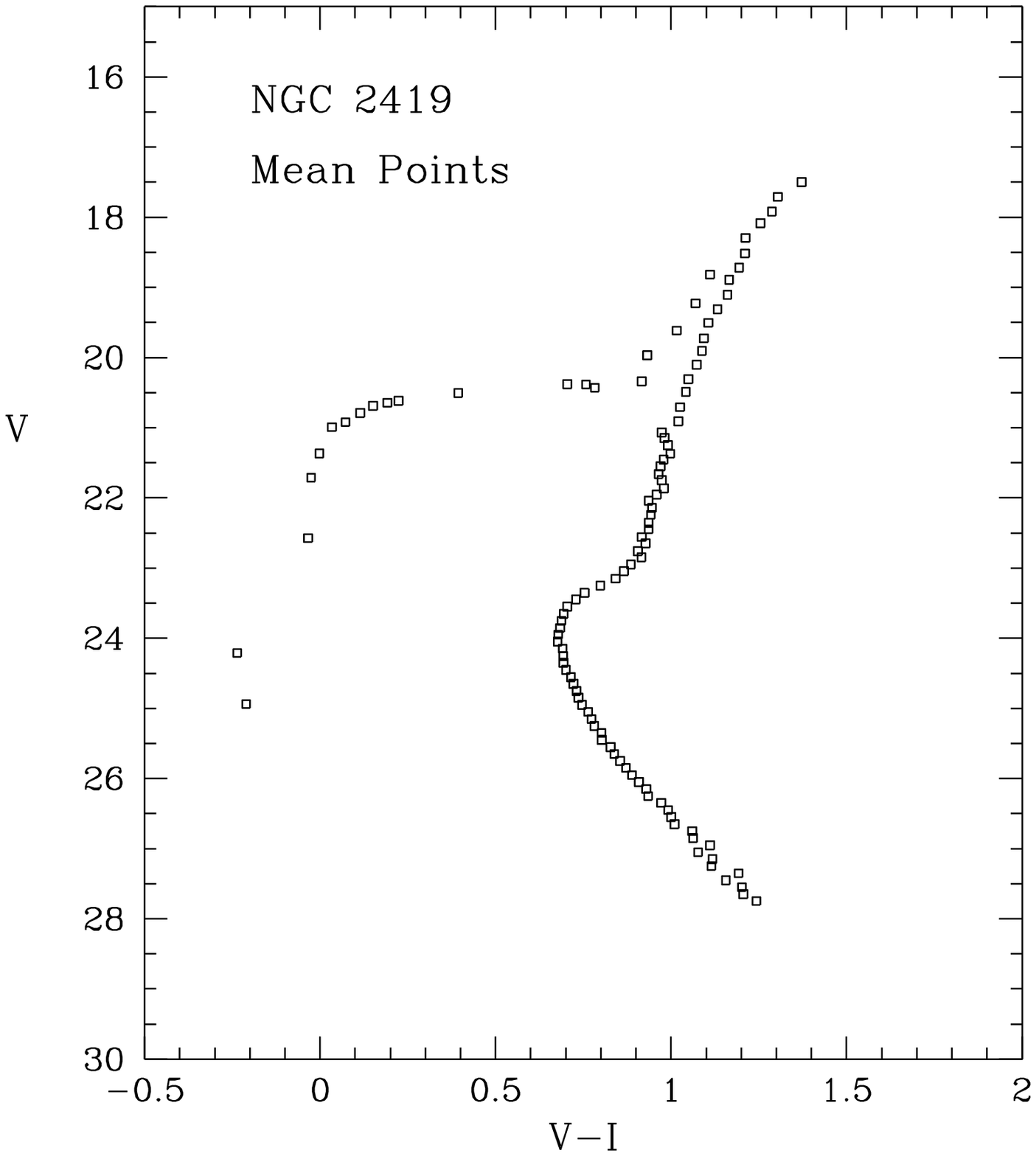]{Fiducial mean points for the color-magnitude
diagram, constructed as described in the text.
For most of the points, the internal error bars are smaller than the 
plotted symbol size.  \label{fig2}}

\figcaption[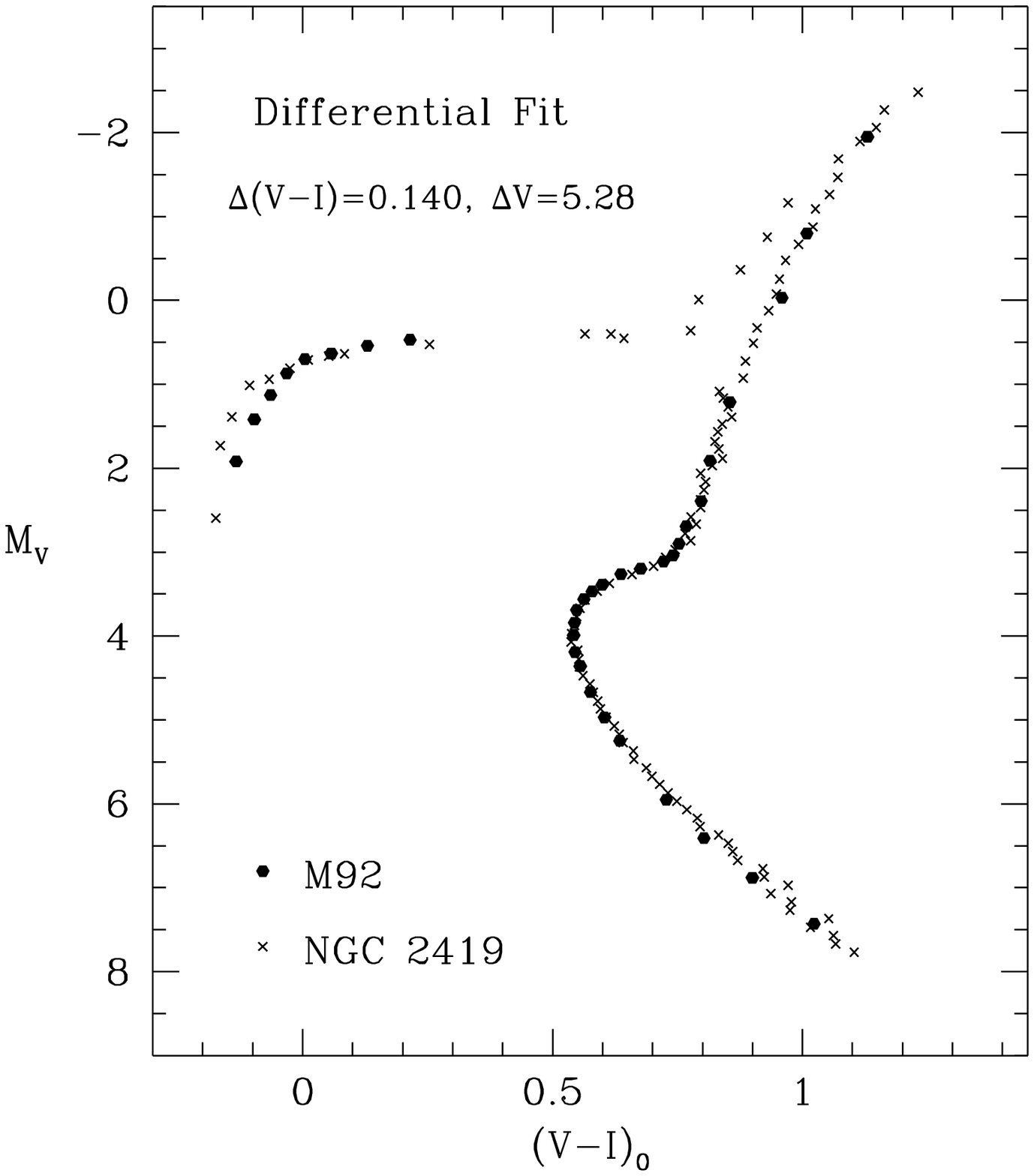]{Differential CMD fit between NGC 2419 (from the present
work) and the similarly metal-poor cluster M92 (from Johnson \& Bolte 1997).
The CMD for NGC 2419 has been shifted brightward and blueward
by the amounts $\Delta V, \Delta(V-I)$ shown in the figure.
Quantitative analysis (see text) shows that the two clusters have the
same age to less than 1 Gyr.  The absolute scales ($M_V, (V-I)_0$) on
the graph have been set by adopting the reddening and distance
modulus for M92 derived later in the text (see Figures 5 and 6), though these absolute
calibrations do not affect the differential fit itself. \label{fig3}}

\figcaption[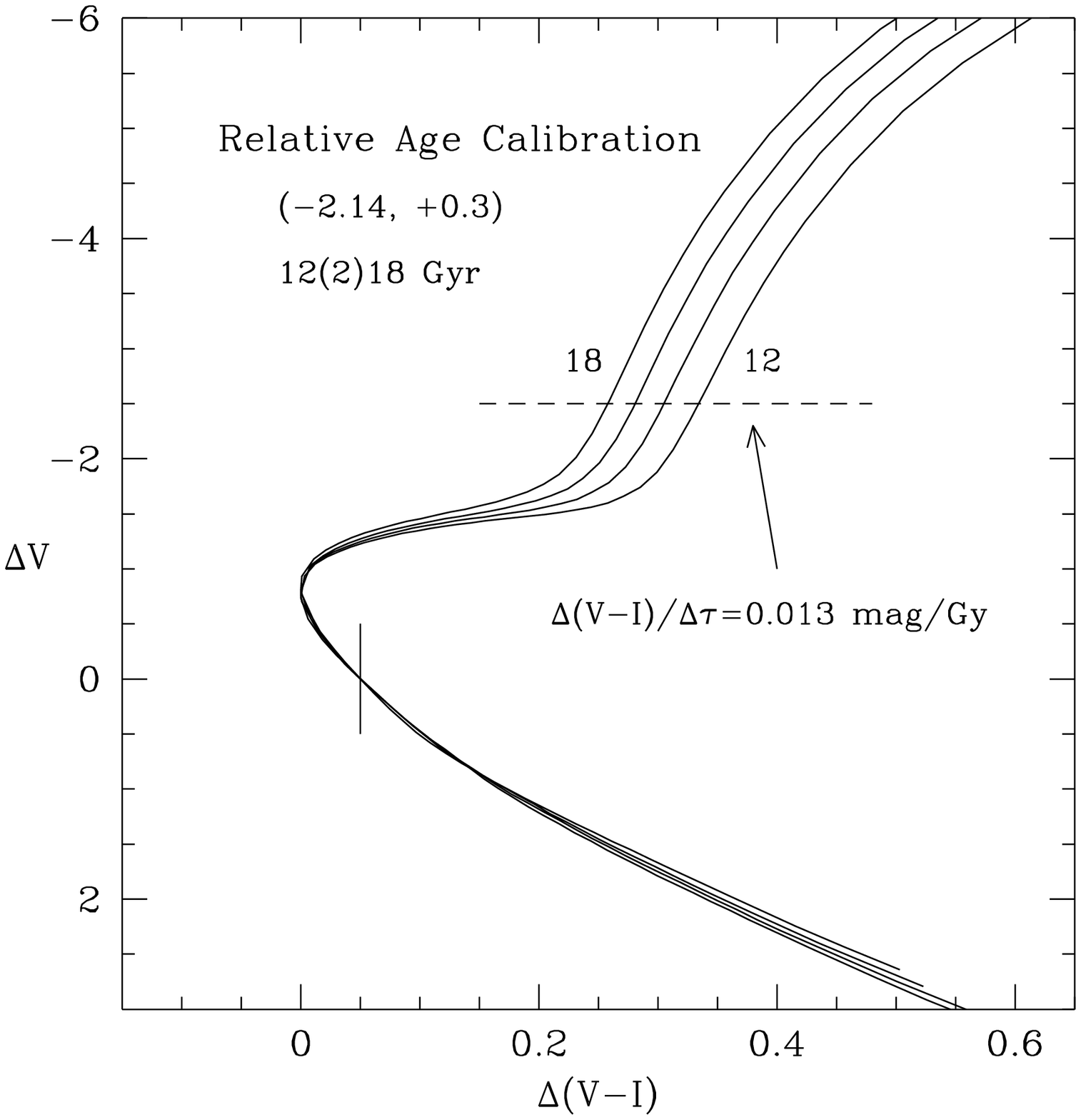]{Model isochrones in $(M_V, V-I)$, derived as described in
\S4 of the text.  The chemical composition of the models is
$Y = 0.235$, [Fe/H] = $-2.14$, and [$\alpha$/Fe] = 0.3.
Isochrone lines are plotted for four different ages (12, 14, 16, 18 Gyr),
shifted arbitrarily so that their main sequence lines
coincide at a point $\Delta(V-I) = 0.05$ mag redder than the turnoff point.
The color difference between the MSTO and the giant branch then changes
with age at the rate shown on the figure.
 \label{fig4}}

\figcaption[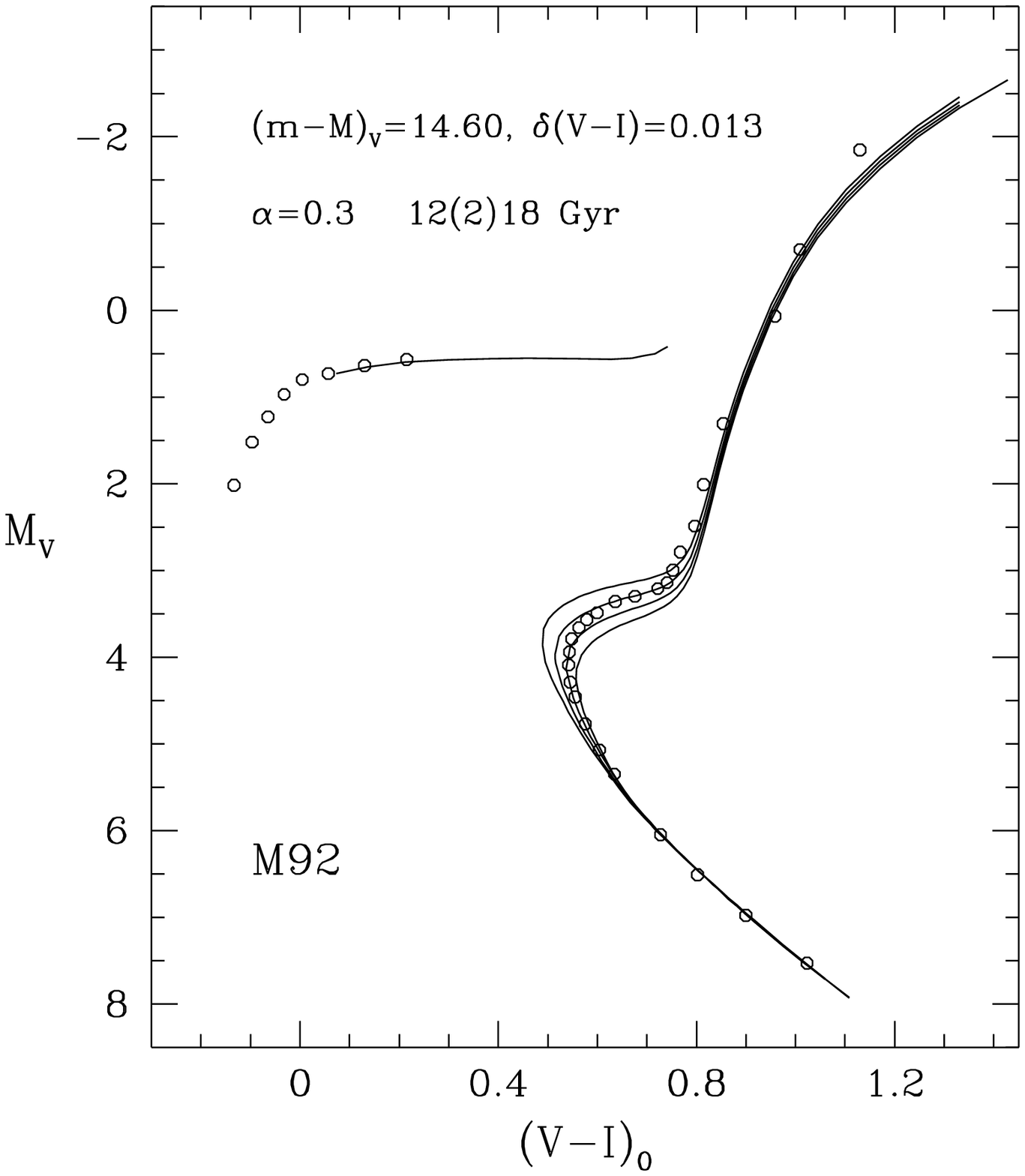]{Isochrone fit and absolute age estimation for M92. 
The adopted isochrones, as in Fig.~4 above, have abundances
$Y = 0.235$, [Fe/H] = $-2.14$, and [$\alpha$/Fe] = 0.3, and
ages of 12, 14, 16, and 18 Gyr.  The fiducial sequences for the cluster have
been shifted by the amounts $\delta(V-I), (m-M)_V$ listed in order
to superimpose them on the isochrones. The resulting RR Lyrae luminosity
is $M_V(HB) = 0.55$, and the best-fit age is $15 \pm 1$ Gyr.  
 \label{fig5}}

\figcaption[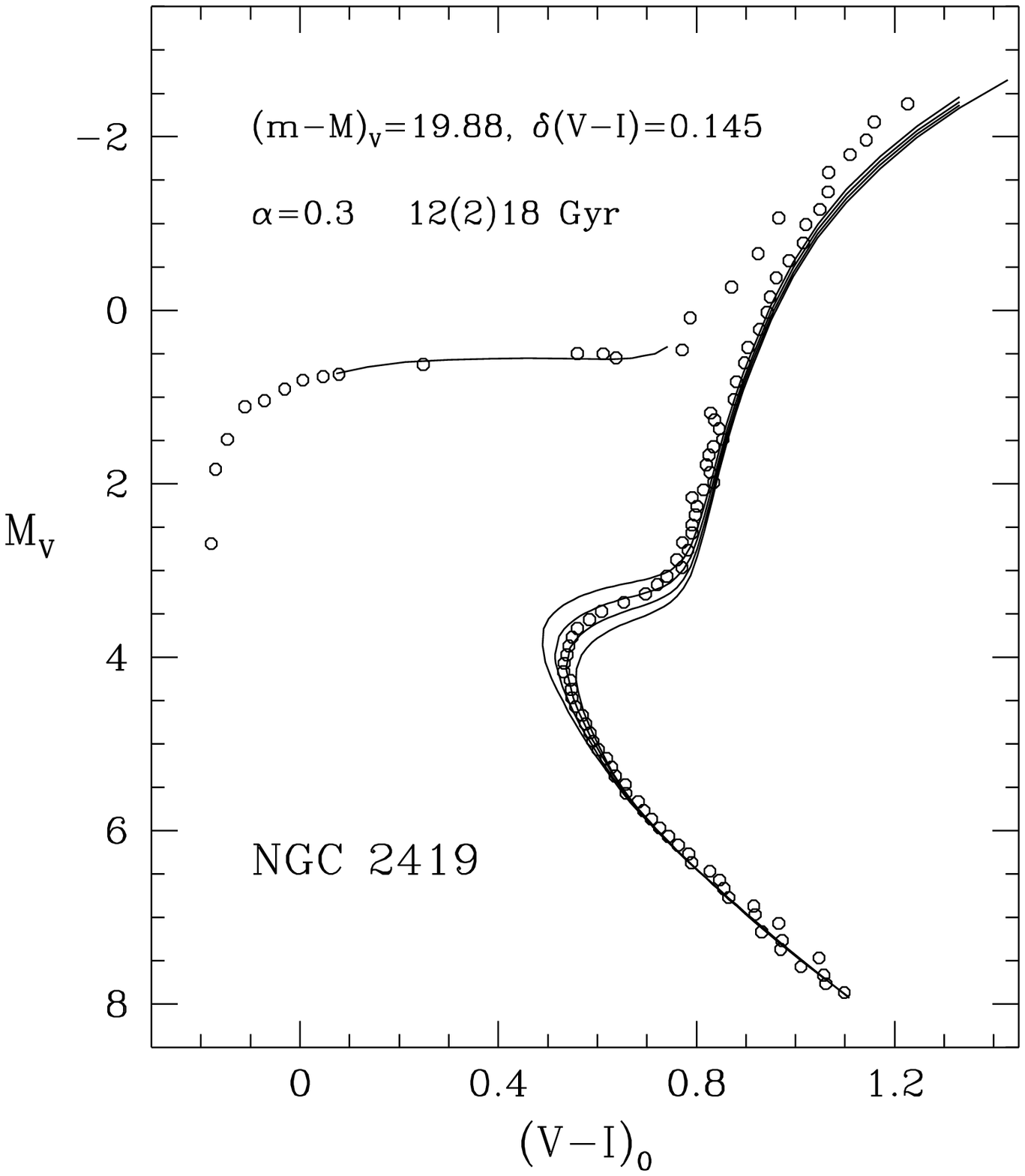]{Isochrone fit to NGC 2419, with the same
stellar models as in the previous figure.  The resulting RR Lyrae luminosity
is $M_V(HB) = 0.57$; the best-fit age is 15 Gyr.  As in Fig.~5, the fit in
this graph optimizes the fit of the isochrones to all parts of the cluster CMD. 
\label{fig6}}

\figcaption[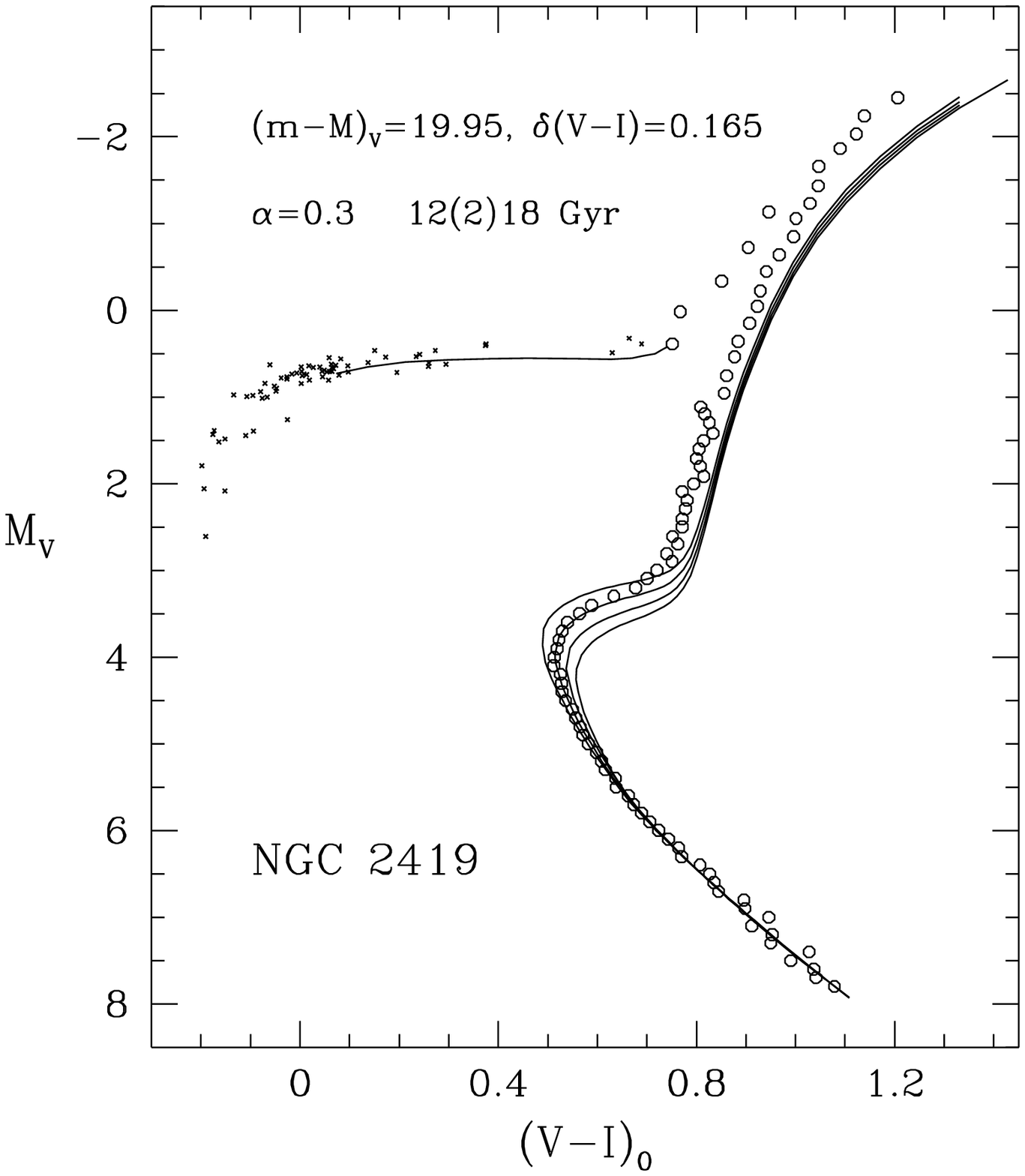]{Isochrone fit to NGC 2419, for a distance modulus
and reddening that optimize the fit strictly to the ZAHB and unevolved main sequence.
Here the horizontal-branch stars are plotted individually (small crosses) to show
the location of the ZAHB line along the lower envelope of the observed HB.
Open circles represent the mean points for the other parts of the CMD.
See \S4.3 for discussion.  \label{fig7}}

\figcaption[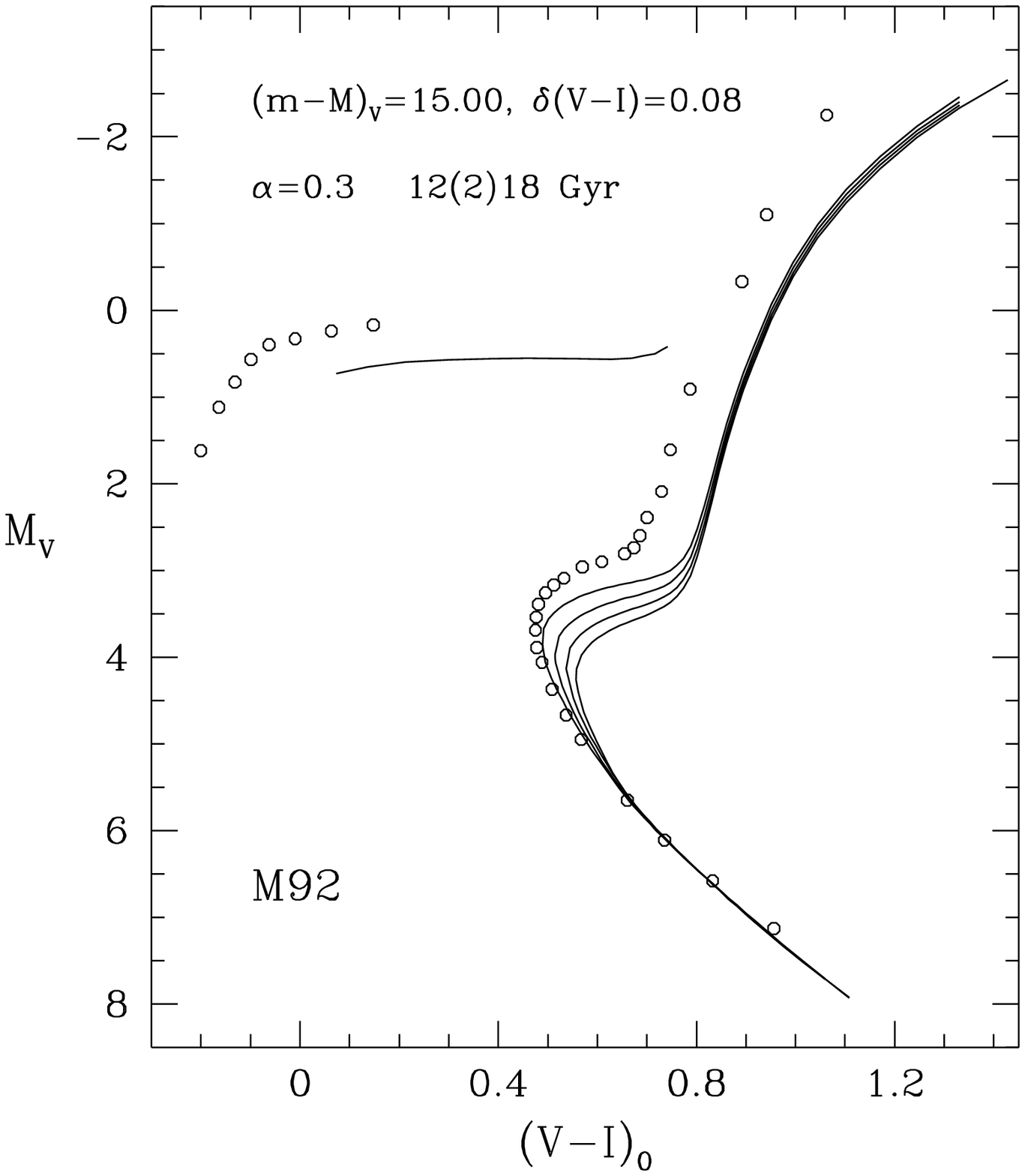]{Isochrone fit to M92.  The isochrone lines are
the same as in the previous figures, but the assumed distance scale is
$M_V(HB) = 0.15$, following the \protect\cite{rei97} analysis of the
{\it Hipparcos} subdwarf parallaxes.  The cluster HB is placed at
$M_V = 0.15$, and the fiducial sequences are then shifted horizontally
until the main sequence points fall in line with the isochrone ZAMS.
The deduced age is 10 Gyr or less.  The deduced `reddening' of M92 in this
case would be $\delta(V-I) = 0.08$ mag.  See \S4.3 of the text for discussion.
 \label{fig8}}

\figcaption[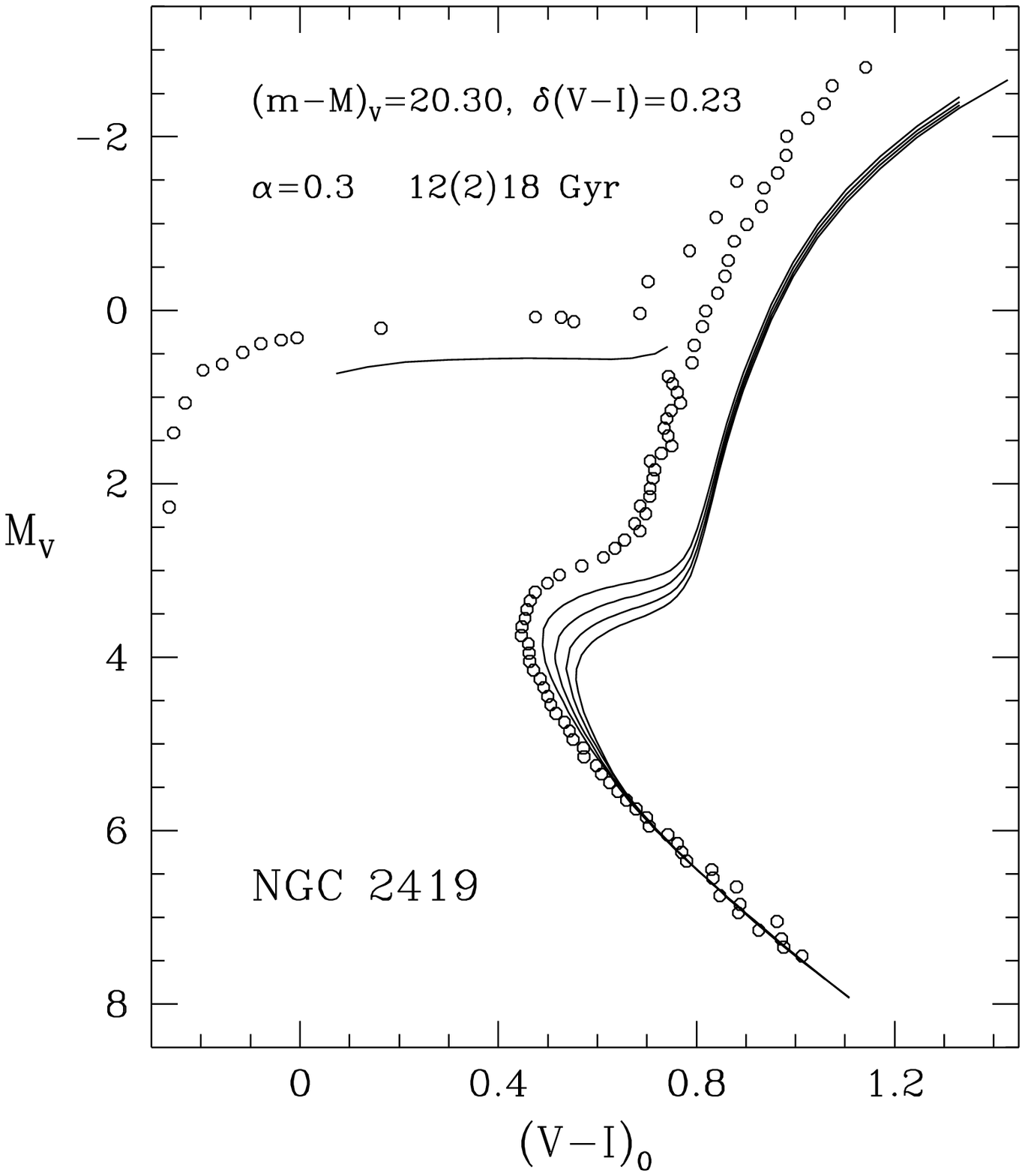]{Isochrone fit to NGC 2419, assuming 
$M_V(HB) = 0.15$ as in the previous figure.  \label{fig9}}

\clearpage
\begin{figure}
\epsscale{1.0}
\plotone{Harris.fig1.ps}
\end{figure}
 
\clearpage
\begin{figure}
\epsscale{1.0}
\plotone{Harris.fig2.ps}
\end{figure}

\clearpage
\begin{figure}
\epsscale{1.0}
\plotone{Harris.fig3.ps}
\end{figure}

\clearpage
\begin{figure}
\epsscale{1.0}
\plotone{Harris.fig4.ps}
\end{figure}

\clearpage
\begin{figure}
\epsscale{1.0}
\plotone{Harris.fig5.ps}
\end{figure}

\clearpage
\begin{figure}
\epsscale{1.0}
\plotone{Harris.fig6.ps}
\end{figure}

\clearpage
\begin{figure}
\epsscale{1.0}
\plotone{Harris.fig7.ps}
\end{figure}

\clearpage
\begin{figure}
\epsscale{1.0}
\plotone{Harris.fig8.ps}
\end{figure}

\clearpage
\begin{figure}
\epsscale{1.0}
\plotone{Harris.fig9.ps}
\end{figure}


\clearpage 
\begin{thebibliography}{}

\bibitem[Alexander \& Ferguson (1994)]{ale94} Alexander, D.R., 
\& Ferguson, J.W. 1994, \apj, 437, 879

\bibitem[Bahcall \& Pinsonneault 1992]{bah92} Bahcall, J.N., 
\& Pinsonneault, M.H. 1992, Rev.Mod.Phys., 64, 885

\bibitem[Beers \etal\ 1990]{bee90} Beers, T.C., Preston, G.W., Shectman, S.A.,
\& Kage, J.A. 1990, \aj, 100, 849

\bibitem[Bell \etal\ 1979]{bel79} Bell, R.A.,
Dickens, R.J., \& Gustafsson, B. 1979, \apj, 229, 604

\bibitem [Bell \& Gustafsson (1989)]{bel89} Bell, R.A., \& Gustafsson, B. 1989, 
\mnras, 236, 653 

\bibitem[Bell \etal\ (1994)]{bel94} Bell, R.A., Paltoglou, G.,
\& Tripicco, M.J. 1994, \mnras, 268, 771

\bibitem[Bell \& Tripicco (1996)]{bel96} Bell, R.A., \& Tripicco, M.J. 
1996, in Stellar Surface Structure, IAU Symposium 176, edited by
K.G. Strassmeier \& J.L.Linsky (Dordrecht:  Kluwer),  p.~527

\bibitem[Bergbusch \& VandenBerg (1992)]{ber92} Bergbusch, P.A., 
\& VandenBerg, D.A. 1992, \apjs, 81, 163

\bibitem[Bessell (1983)]{bes83} Bessell, M.S. 1983, \pasp, 95, 480 

\bibitem[Bessell (1990)]{bes90} Bessell, M.S. 1990, \pasp, 102, 1181 

\bibitem[Bolte 1989]{Bo89} Bolte,~M. 1989, \aj, 97, 1688

\bibitem[Buonanno \etal\ 1983]{buo83} Buonanno, R., Buscema, G., Corsi, C.E.,
Iannicola, G., Smriglio, F., \& Fusi Pecci, F. 1983, \aaps, 53, 1

\bibitem[Buonanno \etal\ 1990]{Bu+90} Buonanno,~R., Buscema,~G.,
Fusi~Pecci,~F., Richer,~H.~B., \& Fahlman,~G.~G. 1990, \aj, 100, 1811

\bibitem[Buonanno \etal\ 1993]{Bu+93} Buonanno,~R., Corsi,~C.~E.,
Fusi~Pecci,~F., Richer,~H.~B., \& Fahlman,~G.~G. 1993, \aj, 105, 184

\bibitem[Carney (1996)]{car96} Carney, B.~W. 1996, \pasp, 108, 900

\bibitem[Carney \etal\ (1992a)]{car92a}  Carney, B.W., Storm, J.,
\& Jones, R.V. 1992a, \apj, 386, 663
 
\bibitem[Carney \etal\ 1992b]{car92b} Carney, B.W., Storm, J., Trammell, S.R.,
\& Jones, R.V. 1992b, \pasp, 104, 44

\bibitem[Castellani \etal\ (1997)]{cas97} Castellani, V., Ciacio, F.,
Degl'Innocenti, S., \& Fiorentini, G. 1997, preprint astro-ph/9705035

\bibitem[Chaboyer \etal\ 1996a]{cha96} Chaboyer, B.,
Demarque, P., \& Sarajedini, A. 1996a, \apj, 459, 558

\bibitem[Chaboyer \etal\ (1996b)]{chab96} Chaboyer, B., Demarque, P.,
Kernan, P.J., Krauss, L.M., \& Sarajedini, A. 1996b, \mnras, 283, 683

\bibitem[Christian \& Heasley (1988)]{chr88} Christian, C.~A., \& 
Heasley, J.~N. 1988, \aj, 95, 1422

\bibitem[Cohen \& Matthews 1992]{coh92b} Cohen, J.G., \& Matthews, K.
1992, \pasp, 104, 1205 

\bibitem[Da Costa \& Armandroff 1995]{dac95} Da Costa, G.S., \& Armandroff, 
T.E. 1995, \aj, 109, 2533

\bibitem[de Boer \etal\ 1997]{deb97} de Boer, K.S., Tucholke, H.-J.,
\& Schmidt, J.H.K. 1997, \aap, 317, L23

\bibitem[Dorman 1992]{dor92} Dorman, B. 1992, \apjs, 81, 221

\bibitem[Dreiling \& Bell (1981)]{dre81} Dreiling, L.A., \& Bell, R.A. 
1981, \apj, 241, 737 

\bibitem[Dufour (1984)]{duf84} Dufour, R.J. 1984, in Structure and Evolution of the
Magellanic Clouds, IAU Symposium 108, edited by S. van den Bergh \&
K.S. de Boer (Dordrecht: Reidel), p.~353

\bibitem[Durrell \& Harris 1993]{DH93} Durrell,~P.~R. \& Harris,~W.~E.
1993, \aj, 105, 1420

\bibitem[Edvardsson \& Bell (1989)]{edv89} Edvardsson, B., \& 
Bell, R.A. 1989, \mnras, 238, 1121

\bibitem[Eggen \etal\ 1962]{ELS} Eggen,~O.~J.,
Lynden-Bell,~D., \& Sandage,~A. 1962, \apj, 136, 748 (ELS)

\bibitem[Feast \& Catchpole 1997]{fea97} Feast, M.W., \& Catchpole, R.M 1997,
\mnras, in press

\bibitem[Fusi Pecci \etal\ 1995]{fus95} Fusi Pecci, F., Bellazzini, M.,
Cacciari, C., \& Ferraro, F.R. 1995, \aj, 110, 1664

\bibitem[Fusi Pecci \etal\ 1996]{fus96} Fusi Pecci, F. 1996, \etal,
\aj, 112, 1461

\bibitem[Gratton \etal\ (1997)]{gra97} Gratton, R.G., Fusi Pecci, F.,
Carretta, E., Clementini, G., Corsi, C., \& Lattanzi, M. 1997, \apj,
submitted (preprint astro-ph/9704150)

\bibitem[Gratton \& Ortolani 1988]{GO88} Gratton,~R.~G., \&
Ortolani,~S. 1988, A\&AS, 73, 137

\bibitem[Green \& Norris 1990]{GN90} Green,~E.~M., \& Norris,~J.~E.
1990, \apjl, 353, L17

\bibitem[Gustafsson \etal\ 1975]{gus75} Gustafsson, B., Bell, R.A.,
Eriksson, K., \& Nordlund,  1975, \aap, 42, 407

\bibitem[Harris 1996]{har96} Harris, W.E. 1996, \aj, 112, 1487

\bibitem[Harris \& Pudritz 1994]{har94} Harris, W.E., \& Pudritz, R.E.
1994, \apj, 429, 177

\bibitem[Harris \etal\ (1995)]{har95} Harris, W.E., Stetson, P.B.,
Bolte, M., Bell, R., Bond, H., Fahlman, G.G., Hesser, J.E., McClure, R.D.,
Richer, H., VandenBerg, D.A., \& van den Bergh, S. 1995, Calibrating Hubble
Space Telescope: Post Servicing Mission, edited by A.Koratkar \& C.Leitherer
(Baltimore:  StScI), p.~283

\bibitem[Hayes (1985)]{hay85} Hayes, D.S. 1985, in Calibration of Fundamental
Stellar Quantities, IAU Symposium No.~111, edited by D.S.Hayes,
L.E.Pasinetti, \& A.G.D.Philip (Dordrecht: Reidel), p.~225

\bibitem[Hesser (1995)]{hes95} Hesser, J.E., in Stellar Populations,
IAU Symposium No.~164, edited by P.C.~van der Kruit \& G.Gilmore
(Dordrecht:  Reidel), p.~51

\bibitem[Hesser (1996a)]{hes96a} Hesser, J.E., Stetson, P.B., Harris, W.E.,
Bolte, M., Smecker-Hane, T.A., VandenBerg, D.A., Bell, R.A., Bond, H.E.,
van den Bergh, S., McClure, R.D., Fahlman, G.G., \& Richer, H.B. 1996a,
J.Korean Astron.Soc., 29, S111

\bibitem[Hesser (1996b)]{hes96b} Hesser, J.E., Stetson, P.B., McClure, R.D., 
van den Bergh, S., Harris, W.E., Bolte, M., VandenBerg, D.A., Bond, H.E.,
Fahlman, G.G., Richer, H.B., \& Bell, R.A. 1996b, \baas, 28, 1363

\bibitem[Holtzman \etal\ (1995)]{hol95} Holtzman, J.A., Burrows, C.J., 
Casertano, S., Hester, J.J., Trauger, J.T., Watson, A.M., \& Worthey, G. 1995, 
\pasp, 107, 1065

\bibitem[Ibata \etal\ 1997]{iba97} Ibata, R.A., Wyse, R.F.G., Gilmore, G.,
Irwin, M.J., \& Suntzeff, N.B. 1997, \aj, 113, 634

\bibitem[Johnson \& Bolte (1997)]{bol97} Johnson, J., \& Bolte, M. 1997, in preparation

\bibitem[Johnson \etal\ 1996]{joh96} Johnson, K.V., Hernquist, L.,
\& Bolte, M. 1996, \apj, 465, 278

\bibitem[Kauffmann \etal\ 1993]{kau93} Kauffmann, G.,
White, S.D.M., \& Guideroni, B. 1993, \mnras, 264, 201

\bibitem[Kraft \etal\ 1997]{kra97} Kraft, R.P., Sneden, C., Smith, G.H.,
Shetrone, M.D., Langer, G.E., \& Pilachowski, C.A. 1997, \aj, 113, 279

\bibitem[Krishna Swamy (1966)]{kri66} Krishna Swamy, K.S. 1966, \apj, 145, 174

\bibitem[Layden \etal\ 1996]{lay96} Layden, A.C., Hanson, R.B., Hawley, S.L.,
Klemola, A.R., \& Hanley, C.J. 1996, \aj, 112, 2110

\bibitem[Lee (1993)]{lee93} Lee, Y.-W. 1993, in The Globular Cluster-Galaxy
Connection, ed.\ G.~Smith and J.~Brodie, ASP Conf.\ Ser., 48, 142

\bibitem[Lee \etal\ 1990]{lee90} Lee, Y.-W., Demarque, P.,
\& Zinn, R. 1990, \apj, 350, 155

\bibitem[Lee \etal\ 1994]{lee94} Lee, Y.-W., Demarque, P.,
\& Zinn, R. 1994, \apj, 423, 248

\bibitem[Lynden-Bell \& Lynden-Bell 1995]{lyn95} Lynden-Bell, D., \& Lynden-Bell, R.M.
1995, \mnras, 275, 429

\bibitem[Magain 1989]{mag89} Magain, P. 1989, \aap, 209, 211

\bibitem[Majewski 1994]{maj94} Majewski, S.R. 1994, \apjl, 431, L17

\bibitem[McLaughlin \& Pudritz 1996]{mcl96} McLaughlin, D.E., \& Pudritz,
R.E. 1996, \apj, 457, 578

\bibitem[Peterson \etal\ 1990]{pet90} Peterson, R.C., 
Kurucz, R.L., \& Carney, B.W. 1990, \apj, 350, 173

\bibitem[Pilachowski \& Armandroff 1996]{pil96} Pilachowski, C.A., \&
Armandroff, T.E. 1996, \aj, 111, 1175

\bibitem[Pont \etal\ (1997)]{pon97} Pont, F., Mayor, M., Turon, C.,
\& VandenBerg, D.A. 1997, \aap, submitted

\bibitem[Proffitt \& VandenBerg (1991)]{pro91} Proffitt, C.R., 
\& VandenBerg, D.A. 1991, \apjs, 77, 473

\bibitem[Racine \& Harris (1975)]{rac75} Racine, R., \& Harris, W.E.
1975, \apj, 196, 413

\bibitem[Reid (1997)]{rei97} Reid, I.N. 1997, \aj, in press

\bibitem[Renzini \etal\ 1996]{ren96} Renzini, A., Bragaglia, A., Ferraro, F.R.,
Gilmozzi, R., Ortolani, S., Holberg, J.B., Liebert, J., Wesemael, F.,
\& Bohlin, R.C. 1996, \apjl, 465, L23

\bibitem[Richer \etal\ 1996]{ric96} Richer, H.B., Harris, W.E., Fahlman, G.G.,
Bell, R.A., Bond, H.E., Hesser, J.E., Holland, S., Pryor, C., Stetson, P.B.,
VandenBerg, D.A., \& van den Bergh, S. 1996, \apj, 463, 602

\bibitem[Rodgers \& Paltoglou 1984]{rog84} Rodgers, A.W., \& Paltoglou, G.
1984, \apjl, 283, L5

\bibitem[Rogers \& Iglesias (1992)]{rog92} Rogers, F.J., \& Iglesias, 
C.A. 1992, \apj, 401, 361

\bibitem[Saha \etal\ 1992]{sah92} Saha, A., Freedman, W.~L., Hoessel, J.~G.,
\& Mossman, A.~E. 1992, \aj, 104, 1072

\bibitem[Salaris \etal\ 1993]{sal93} Salaris, M., Chieffi, A.,
\& Straniero, O. 1993, \apj, 414, 580

\bibitem[Salaris \etal\ (1997)]{sal97} Salaris, M., Degl'Innocenti, S.,
\& Weiss, A. 1997, \apj, 479, 665

\bibitem[Salaris \& Weiss 1997]{wei97} Salaris, M., \& Weiss, A. 1997,
preprint astro-ph/9704238

\bibitem[Sandage 1969]{san69} Sandage, A. 1969, \apj, 157, 515

\bibitem[Sandage (1990)]{Sa90} Sandage,~A. 1990, \jrasc, 84, 70

\bibitem[Sandquist \etal\ 1996]{san96} Sandquist, E.L., Bolte, M,
Stetson, P.B., \& Hesser, J.E. 1996, \apj, 470, 910

\bibitem[Sarajedini 1997]{sar97} Sarajedini, A. 1997, \aj, 113, 683

\bibitem[Sarajedini \& Demarque 1990]{sar90} Sarajedini,~A., \&
Demarque,~P. 1990, \apj, 365, 219

\bibitem[Sarajedini \& King 1989]{sar89} Sarajedini, A., \&
King, C.R. 1989, \aj, 98, 1624

\bibitem[Searle \& Zinn (1978)]{SZ78} Searle,~L., \& Zinn,~R. 1978,
\apj, 225, 357

\bibitem[Shetrone 1996]{she96} Shetrone, M.D. 1996, \aj, 112, 1517

\bibitem[Silk \& Wyse 1993]{sil93} Silk, J., \& Wyse, R.F.G. 1993,
Phys.Rep., 231, 293

\bibitem[Sneden \etal\ 1991]{sne91} Sneden, C., Kraft, R.~P., 
Prosser, C.~F., \& Langer, G.~E. 1991, \aj, 102, 2001

\bibitem[Stetson \etal\ 1989]{pbs+89} Stetson,~P.~B., VandenBerg,~D.~A.,
Bolte,~M., Hesser,~J.~E., \& Smith,~G.~H. 1989, \aj, 97, 1360

\bibitem[Stetson \etal\ 1996]{ste96} Stetson, P.B.,
VandenBerg, D.A., \& Bolte, M. 1996, \pasp, 108, 560 (SVB)

\bibitem[Stetson \& West 1994]{ste94} Stetson, P.B., \& West, M.J. 1994,
\pasp, 106, 726

\bibitem[Stetson \etal\ (1997)]{ste97} Stetson, P.B., \etal\ 1997,
in preparation

\bibitem[Sturch (1966)]{stu66} Sturch, C.R. 1966, \apj, 143, 774

\bibitem[Suntzeff \etal\ 1988]{sun88}
Suntzeff, N.B., Kraft, R.P., and Kinman, T.D. 1988, \aj, 95, 91

\bibitem[Tripicco \& Bell (1995)]{tri95} Tripicco, M.J., \& Bell, R.A. 
1995, \aj, 110, 3035 

\bibitem[van den Bergh (1993)]{vdb93} van den Bergh, S. 1994, \apj, 411, 178

\bibitem[van den Bergh 1994]{van94} van den Bergh, S. 1994, \aj, 108, 2145

\bibitem[van den Bergh 1995a]{van95a} van den Bergh, S. 1995a, \aj, 110, 1171

\bibitem[van den Bergh 1995b]{van95b} van den Bergh, S. 1995b, \apj, 446, 39

\bibitem[van den Bergh \etal\ 1991]{van91} van den Bergh, S.,
Morbey, C., \& Pazder, J. 1991, \apj, 375, 594

\bibitem[VandenBerg 1992]{van92} VandenBerg, D.A. 1992, \apj, 391, 685

\bibitem[VandenBerg (1997)]{vdb97} VandenBerg, D.A. 1997, in Fundamental
Stellar Properties:  The Interaction Between Theory and Observation,
IAU Symposium 189, edited by T.R.Bedding, A.J.Booth, \& J.Davis (Dordrecht:
Kluwer), in press

\bibitem[VandenBerg \& Bell (1985)]{van85} VandenBerg, D.A., \& Bell,R.A. 1985,
\apjs, 58, 561 (VB85)

\bibitem[VandenBerg \etal\ 1990]{VBS90} VandenBerg,~D.~A.,
Bolte,~M., \& Stetson,~P.~B. 1990, \aj, 100, 445 (VBS90)

\bibitem[VandenBerg \etal\ 1996]{van96} VandenBerg,~D.~A.,
Bolte,~M., \& Stetson, P.B. 1996, \araa, 34, 461 (VBS96)

\bibitem[VandenBerg \etal\ (1997)]{van97} VandenBerg, D.~A., Swenson, F.J., 
Rogers, F.J., Iglesias, C.A., \& Alexander, D.R. 1997, in preparation

\bibitem[Walker 1992]{arw92} Walker,~A.~R. 1992, \apjl, 390, L81

\bibitem[Wheeler \etal\ 1989]{whe89} Wheeler, J.C.,
Sneden, C., \& Truran, J.W. 1989, \araa, 27, 279

\bibitem[Zhao \& Magain (1990)]{zha90} Zhao, G., \& Magain, P. 1990, \aap,
238, 242

\bibitem[Zinn 1985]{zin85} Zinn, R. 1985, \apj, 293, 424

\bibitem[Zinn (1993)]{Zi93} Zinn,~R. 1993, in The Globular Cluster-Galaxy
Connection, ed.\ G.~Smith and J.~Brodie, ASP Conf.\ Ser., 48, 38

\end{thebibliography}
\end{document}